\documentclass[11pt]{article}
\usepackage{geometry}                
\usepackage{graphicx}
\usepackage{amssymb}
\usepackage{mathtools}
\usepackage{verbatim}
\usepackage{color}
\usepackage{slashed}
\usepackage{bbold}
\usepackage{fullpage}
\usepackage{float}
\usepackage[colorlinks]{hyperref}
\usepackage[font=footnotesize,labelfont=bf]{caption}
\usepackage{mathabx}
\usepackage{multirow}

\newcommand{\beq}{\begin{equation}}
\newcommand{\eeq}{\end{equation}}
\newcommand{\bmat}{\begin{pmatrix}}
\newcommand{\emat}{\end{pmatrix}}
\newcommand{\bal}{\begin{align}}
\newcommand{\eal}{\end{align}}

\newcommand{\MS}{\overline{MS}}
\newcommand{\DR}{\overline{DR}}

\begin{document}

\def\simgt{\mathrel{\lower2.5pt\vbox{\lineskip=0pt\baselineskip=0pt
           \hbox{$>$}\hbox{$\sim$}}}}
\def\simlt{\mathrel{\lower2.5pt\vbox{\lineskip=0pt\baselineskip=0pt
           \hbox{$<$}\hbox{$\sim$}}}}

\begin{titlepage}

\rightline{SLAC-PUB-17292}
\rightline{UMN--TH--3722/18, FTPI--MINN--18/12}
\vskip 0.2in

\begin{center}

{\LARGE \bf New Weak-Scale Physics from SO(10) with \\High-Scale Supersymmetry}
\vspace{0.8cm}

\small
{\bf Sebastian~A.~R.~Ellis$^a$},
{\bf Tony Gherghetta$^{b}$},
{\bf Kunio Kaneta$^{b,c}$},
and {\bf Keith~A.~Olive$^{b,c}$}
\normalsize

\vspace{.5cm}
$^a$\textit{SLAC National Accelerator Laboratory, 2575 Sand Hill Road, Menlo Park, CA 94025, USA}\\
${}^b$\textit{School of Physics and Astronomy, University of Minnesota, Minneapolis, MN 55455, USA}\\
${}^c$\textit{William I.~Fine Theoretical Physics Institute, University of Minnesota, Minneapolis, MN 55455, USA}
\vspace{1 cm}

\begin{abstract}
Gauge coupling unification and the stability of the Higgs vacuum are among two of the cherished features of low-energy supersymmetric models. 
Putting aside questions of naturalness, supersymmetry might only be realised in nature at very high energy scales. 
If this is the case, the preservation of gauge coupling unification and the stability of the Higgs vacuum would certainly require new physics, but it need not necessarily be at  weak scale energies. 
New physics near the unification scale could in principle ensure Grand Unification, while new physics below $\mu \sim 10^{10}$ GeV could ensure the stability of the Higgs vacuum. 
Surprisingly however, we find that in the context of a supersymmetric SO(10) Grand Unified Theory, gauge coupling unification and the Higgs vacuum stability, when taken in conjunction with existing phenomenological constraints, require the presence of $\mathcal{O}$(TeV)-scale physics. 
This weak-scale physics takes the form of a complex scalar SU(2)$_L$ triplet with zero hypercharge, originating from the {\bf 210} of SO(10).

\end{abstract}
\end{center}

\end{titlepage}

\section{Introduction}

One of the most conspicuous null results of the LHC run I and run II so far, has been the lack of discovery of supersymmetry (SUSY) \cite{nosusy} near the electroweak scale.
There are at least three possible implications of this result. First, perhaps the supersymmetric mass scales are just around the corner at the multi-TeV scale which may or may not be within the reach of the LHC. Second, 
supersymmetry could be broken at a very high energy scale (though below the Grand Unified Theory (GUT) scale), in which case
the supersymmetric particle spectrum would not be directly accessible at the LHC. Finally, it is also possible that supersymmetry is not manifest
below the Planck scale and is an exact symmetry only at the string or Planck scale. In this case, it is unlikely that there are any
experimental consequences of supersymmetry.  While there have been many studies of supersymmetry at the multi-TeV scale, we instead examine the second alternative, namely that of high-scale supersymmetry \cite{hssusy,hssusy2}. This is partly motivated by the possibility that an EeV mass gravitino may provide the correct relic density of dark matter \cite{DMO}
if the supersymmetry breaking scale lies above the inflationary scale, $m_I\simeq 3 \times 10^{13} $ GeV. Thus while high-scale supersymmetry
may still provide a viable dark matter candidate, it is less clear whether or not supersymmetry can still provide
successful gauge coupling unification or the stability of the Higgs vacuum. It goes without saying that the supersymmetric solution to 
the hierarchy problem will not be available. 

Gauge coupling unification can be achieved in non-supersymmetric GUT models such as SO(10) \cite{so10} which break through an intermediate scale 
gauge group \cite{GN2,so10gc,moqz,mnoqz,noz,mnoz}. The running of the gauge couplings can be deflected at the intermediate
scale when new particle degrees of freedom appear. These same SO(10) models can also stabilize the Higgs vacuum \cite{mnoz} and provide for a dark matter candidate if the intermediate scale is broken by a {\bf 126} dimensional representation \cite{mnoqz,noz}.
Gauge coupling unification in supersymmetric models of SO(10) has also been studied extensively
\cite{gcu,AG}. In this case, since supersymmetry alone is sufficient for the focusing of the gauge coupling running \cite{Ellis:1990zq}, care in the construction of 
the GUT model is needed to avoid spoiling the success achieved in the minimal supersymmetric Standard Model (MSSM) if new states
remain light below the GUT scale.  One of the main objectives of this paper is to discuss the implications of the spectrum of states in an SO(10) GUT on obtaining satisfactory high-scale SUSY SO(10) unification.

A viable dark matter candidate is one of the many motivations for low energy supersymmetric models.  Often this candidate is a neutralino
\cite{gold,EHNOS} whose relic density is obtained when thermal annihilations or co-annihilations \cite{gs} freeze-out. However, as the neutralino
mass scale is increased (for example due to LHC lower limits), annihilation and co-annihilation cross sections become too weak
to maintain equilibrium and an excess relic density is left behind. Certainly for neutralino masses in excess of ${\cal O}(10)$ TeV, this thermal
picture breaks down. Alternatively, the gravitino is also an excellent dark matter candidate \cite{pp,nos,EHNOS,oss,eoss5,0404231,stef,buch,rosz,covi}
produced primarily during reheating in processes such as gluon + gluon $\to$ gluino + gravitino.
A combination of limits from big bang nucleosynthesis \cite{bbn} and decays from a next-to-lightest neutralino \cite{myy,kmy,ego}
place an upper bound of $m_{3/2} \lesssim 4$ TeV. However if the SUSY spectrum is pushed to very high scales (above $m_I$),  single gravitino production is kinematically cut off, and the suppressed two gravitino channels dominate
opening a new window for gravitino dark matter with masses between $0.1 - 1000$ EeV \cite{bcdm,DMO,dgmo}. Indeed, in \cite{dgmo}, an integrated 
model of supersymmetry breaking and inflation was constructed with supersymmetry breaking masses, ${\widetilde m} > m_I$. Possible neutrino signatures of this 
model if $R$-parity is not exact were discussed in \cite{dgkmo}. 

As noted above, such a model has little chance in addressing the hierarchy problem, but should be able to still
address other features commonly associated with low energy SUSY. These include obtaining a Higgs mass of 125 GeV,
gauge coupling unification, stability of the Higgs vacuum and dark matter. In \cite{dgmo}, the question of the dark matter
abundance through reheating in a specific inflationary model was addressed and solutions to these other questions were outlined.
Here, we address these issues in more detail in the context of a supersymmetric SO(10) GUT. 

In particular, we consider the so-called minimal supersymmetric GUT based on SO(10) described in detail in \cite{abmsv,bmsv,AG}.
The Higgs sector contains a {\bf 210} to break SO(10), a {\bf 126} and $\bf \overline{126}$ pair to break the intermediate scale gauge group,
and a {\bf 10} to break the Standard Model (SM). Matter fields of each generation are neatly contained in a fundamental {\bf 16}. 
As we will see, in general, vacuum expectation values (VEVs) for the {\bf 210} and {\bf 126} and  $\bf \overline{126}$ occur simultaneously
so that in effect SO(10) is broken directly to the MSSM. Depending on the pattern of VEVs, some states may 
remain light, i.e., below the GUT scale. Since $\widetilde m> m_I$, all $R$-parity $-1$ states (except perhaps the gravitino) also retain masses in excess of $m_I$.  
Indeed, non-negligible supersymmetry breaking effects alter the mass spectra produced in GUT symmetry breaking as we discuss in detail below.

It is often noted that whereas gauge coupling unification is absent in the standard model,
it occurs quite naturally in the MSSM \cite{Ellis:1990zq} with supersymmetric states near the TeV scale. 
However, it has recently been emphasized \cite{Ellis:2015jwa} that unification is also achievable with high-scale supersymmetry. Depending on the GUT gauge group and the superheavy mass spectrum, unification may still 
occur where the mismatch between the low-energy gauge couplings and the GUT (unified) gauge coupling is accounted for by threshold corrections \cite{Hall:1980kf,Langacker:1992rq,Ellis:2015jwa}.  
Indeed for a suitably complicated GUT such as SO(10) with an extensive GUT Higgs structure,
these threshold corrections may in fact be quite significant. 
In the present context of high-scale supersymmetry, we expect SM running of the gauge couplings up to the inflationary scale
which is only slightly below the GUT scale.  Nevertheless, as we will see, gauge coupling unification can still be achieved
when properly taking into account the predicted mass spectrum of superheavy states.  This does depend on the supersymmetry-breaking mechanism, and for simplicity we will assume that all the MSSM superpartners are degenerate in mass at the scale $\widetilde m$ to highlight the effect of the threshold corrections at the GUT scale, while the gravitino mass is implicitly assumed to be $0.1 - 1000$ EeV.
Besides gauge coupling unification the resolution of the Higgs stability question presumably requires some modification to the SM below the scale of $10^{10}$ GeV (where the Higgs quartic coupling
runs negative), and it is quite possible that some component of either the {\bf 210} or {\bf 126} may remain light. 
However as we will argue below, there is only a single candidate for this light state in the minimal SO(10) model.
This state is an SU(2)$_L$ triplet, color singlet with zero hypercharge contained in the {\bf 210}, labelled $S$. 
As will be shown, the threshold corrections for each of the three SM gauge groups, though large, are similar in magnitude and therefore
some focusing of the gauge coupling running, beyond what occurs in the SM, remains necessary.
The state $S$ has a small yet important focusing effect on the running of the gauge couplings
and together with the large threshold effects from GUT states gives rise to precise unification. In contrast, every other charged (or singlet) component of the GUT representations, if light, would negate unification due to the contribution of threshold effects from the light states and the GUT states.

Interestingly, when phenomenological constraints on the $S$ state are taken into account, we find that it should not have too \textit{large} a mass. This is due to the long lifetime of the state, which is such that it contributes a sub-component of the dark matter abundance. Thus, we will show how the SO(10) GUT provides a viable embedding of high-scale supersymmetry, with a stable Higgs vacuum and correct gauge coupling unification, but only as long as there is a TeV-scale particle in the spectrum.

In what follows, 
 we will first go over the minimal field content in section 2, and discuss known solutions for breaking SO(10) while preserving $G_{SM} = {\rm SU(3)}_c\times {\rm SU(2)}_L \times {\rm U(1)}_Y$ in a supersymmetric context. 
There are a number of solutions that break SO(10) to $G_{SM}$ directly, many of which have states much lighter than the GUT scale. 
We discuss the running of the gauge couplings in section 3, and our treatment of threshold corrections in section 4. In section 5, we discuss specific
solutions where gauge coupling unification is achieved and the running of the Higgs quartic coupling is discussed in section 6.
Our conclusions are summarized in section 7.

\section{The minimal GUT field content, interactions, VEVs and masses}
\label{GUTfields.SEC}

We will follow the analysis of \cite{abmsv,bmsv,AG} and include only the following Higgs superfields 
(with SO(10) representation in parentheses):
\beq
\Phi ({\bf 210}); ~~~\Sigma ({\bf 126});~~~\overline{\Sigma} ({\bf \overline{126}});~~~H({\bf 10}) \ .
\eeq
The most general renormalizable superpotential in terms of this field content is:
\beq
W \supset \frac{\mu_\Phi}{4!}\Phi^2 + \frac{\mu_\Sigma}{5!}\Sigma\overline{\Sigma} + \frac{\lambda}{4!} \Phi^3 + \frac{\eta}{4!}\Phi\Sigma\overline{\Sigma} + \mu_H H^2 + \frac{1}{4!} \Phi H (\alpha \Sigma + \overline{\alpha}\overline{\Sigma}) \ ,
\label{eq:superpot}
\eeq
where $\mu_\Phi,\mu_\Sigma, \mu_H$ are mass parameters and $\lambda, \eta,\alpha, \overline{\alpha}$ are dimensionless couplings.
In addition, the theory contains three generations of matter representations $\Psi$({\bf 16}) which couple to the Higgs fields $H$ and $\overline{\Sigma}$.  
Given the relatively large representations we are forced to utilize, it is useful to decompose them down to smaller representations given in terms
of the SU(2)$_L \times$ SU(2)$_R \times$ SU(4) subgroup of of SO(10). These are:
\begin{align}
\label{10rep}
{\bf 10} = (1,1,6) + (2,2,1)\,, \\
{\bf 126} = (1,3, \overline{10}) + (3,1,10) + (1,1,6) + (2,2,15)\,, \\
{\bf \overline{126}} = (1,3, 10) + (3,1,\overline{10}) + (1,1,6) + (2,2,15)\,, \\
\label{210rep}
{\bf 210} = (1,1,15) + (1,1,1) + (1,3,15) + (3,1,15) + (2,2,6) + (2,2,10) + (2,2,\overline{10}) \, .
\end{align}
We further recall the SU(4) decomposition in terms of its ${\rm SU(3)}_c \times {\rm U(1)}_{B-L}$ subgroup: {\bf 6} = 3(2/3) + $\overline{3}$(-2/3), {\bf 10} = 6(-2/3) + 3(2/3) + 1(2), and {\bf 15} = 8(0) + 3(-4/3) + $\overline{3}$(4/3) + 1(0).

Since only MSSM singlets can obtain GUT scale VEVs, there is a limited number of fields which are allowed to obtain VEVs and break SO(10). These are defined as
\begin{align}
v_{1,1,1} = \langle \Phi(1,1,1)\rangle;~~~v_{1,1,15}=\langle\Phi(1,1,15)\rangle;~~~v_{1,3,15} = \langle \Phi(1,3,15)\rangle;~ \\
\sigma_{1,3,\overline{10}} = \langle \Sigma(1,3,\overline{10})\rangle;~~~\sigma_{1,3,10} = \langle \overline{\Sigma}(1,3,10)\rangle \ ,
\end{align}
which means the superpotential for the VEVs can be written as
\begin{align}
W &\supset ~\mu_\Phi\left( v_{1,1,1}^2 + 3 v_{1,1,15}^2 + 6 v_{1,3,15}^2\right) + 2 \lambda\left( v_{1,1,15}^3 + 3 v_{1,1,1} v_{1,3,15}^2+6 v_{1,1,15}v_{1,3,15}^2\right) \\
\nonumber&+ \mu_\Sigma\,\sigma_{1,3,\overline{10}}\sigma_{1,3,10} + \eta\, \sigma_{1,3,\overline{10}}\sigma_{1,3,10}\left( v_{1,1,1}+3 v_{1,1,15}-6 v_{1,3,15}\right)  \ .
\end{align}
Imposing the condition of vanishing $D$-terms implies $|\sigma_{1,3,\overline{10}}|=|\sigma_{1,3,10}| $, and imposing the condition of vanishing $F$-terms leads to the following equations:
\begin{align}
2\mu_\Phi v_{1,1,1} + 6 \lambda v_{1,3,15}^2 + \eta \sigma_{1,3,\overline{10}}\sigma_{1,3,10} = 0 \ ,
\\
2\mu_\Phi v_{1,1,15} + 2 \lambda \left( v_{1,1,15}^2 + 2v_{1,3,15}^2\right) + \eta \sigma_{1,3,\overline{10}}\sigma_{1,3,10} = 0 \ ,
\\
2\mu_\Phi v_{1,3,15} + 2 \lambda \left(v_{1,1,1} + 2v_{1,1,15}\right)v_{1,3,15} + \eta \sigma_{1,3,\overline{10}}\sigma_{1,3,10} = 0 \ ,
\\
\sigma_{1,3,\overline{10}}\left( \mu_\Sigma + \eta\left( v_{1,1,1}+3v_{1,1,5}-6v_{1,3,15}\right)\right)= 0 \, .
\end{align}

There are several solutions to this set of equations including the trivial one with all VEVs equal to zero, for which SO(10) is preserved.
Other solutions include the breaking of SO(10) to either SU(5)$\times$U(1), flipped SU(5)$\times$U(1), SU(5), SU(3)$_c \times$ SU(2)$_L \times$ SU(2)$_R \times$ U(1)$_{B-L}$, or  SU(3)$_c \times$ SU(2)$_L \times$ U(1)$_R \times$ U(1)$_{B-L}$.
It is also possible to break SO(10) directly down to the SM. As we will be primarily interested in this class of solutions, we quote the 
general solution to these conditions from \cite{bmsv}:
\begin{align}
v_{1,1,1} = -\frac{\mu_\Phi}{\lambda}\frac{x(1-5x^2)}{(1-x)^2};~~
v_{1,1,15} = -\frac{\mu_\Phi}{\lambda}\frac{(1-2x-x^2)}{(1-x)};~~v_{1,3,15}=-\frac{\mu_\Phi}{\lambda}x;~~\nonumber \\
\sigma_{1,3,\overline{10}}\sigma_{1,3,10} = \frac{2\mu_\Phi^2}{\eta \lambda}\frac{x(1-3x)(1+x^2)}{(1-x)^2};~~
-8x^3+15x^2-14x+3=(x-1)^2\frac{\lambda \mu_\Sigma}{\eta \mu_\Phi} \ .
\label{sol}
\end{align}
Given the parameters $\mu_\Phi, \mu_\Sigma, \lambda$, and $\eta$, the final equation in (\ref{sol}) determines $x$ which in turn determines each of the 5 VEVs. 

Ignoring for the moment the effects of supersymmetry breaking, 
the value of $x$ also determines the mass spectrum of the SM components in $\Phi$, $\Sigma$, and $\overline{\Sigma}$. 
The masses of these states have been determined in \cite{bmsv}. Some of the mass eigenstates reside purely in either $\Phi$ or $\Sigma$, and $\overline{\Sigma}$, while others correspond to mixed states. 
The mass spectrum of the unmixed SM components as a function of $x$, given in Table I of \cite{bmsv}, is reproduced here for convenience in Table \ref{UnmixedScalars.TAB}, with the coupling conventions labelled as in \cite{AG}. The mixed scalar states are given in Table  \ref{MixedScalars.TAB}.
For the fermionic superpartners of the mixed scalar states, gauginos also participate in the mixings in the $J$, $F$, $E$, $X$, and $G$ states.
For instance, the number of $G$ boson states is five ($G_{1-5}$), while that of $G$ fermion states is six ($G_{1-6}$) as the fermionic superpartner of  the $(1,1,0)$ gauge boson should also be counted.
We will use these label conventions to easily distinguish between different possible 
solutions for light states. Note however, that in our notation, we have defined hypercharge as $Y = {T_3}_R + (B-L)/2$ and $Q = {T_3}_L + Y$.
For a generic value of $x$, one would expect that all of the Higgs states (other than the SM Higgs doublet which must be tuned to
a weak scale value), have masses of order the GUT scale. However, when $x$ is a root (or close to a root) of one of the polynomials listed in the 3rd
column of Table \ref{UnmixedScalars.TAB}, that state will be light.
The mass eigenvalues of the mixed states in the absence of supersymmetry breaking can also be found in \cite{bmsv,AG}.
Note that the massless states correspond to the Nambu-Goldstone bosons which combine with the gauge bosons to give them masses\footnote{Note that the expressions given in Table II of Ref.\cite{bmsv} are incorrect.  The values for the masses of the mixed states are obtained by diagonalizing the matrices given in \cite{AG}.}.

\begin{table}
\centering
\begin{tabular}{| c | c | c | c | c | c | }
\hline\hline
\textbf{Field} &  $(SU(3)_c,SU(2)_L,U(1)_Y)$ & \textbf{Fermion Mass /}$\mu_\Phi$ &$\ell_1$& $\ell_2$ & $\ell_3$\\
\hline 
\hline
\rule{0pt}{3ex}
$\Phi$ & $(3,1,5/3)$+h.c. (\textit{I}) & $8x(2x-1)/(x-1)^2$ & $10$ & 0 & 1 \\ 
\rule{0pt}{3ex}
           & $(8,1,1)$+h.c. (\textit{Z}) & $4(x^2-3x+1+3x^3)/(x-1)^2$ &$\frac{48}{5}$ & 0 & 6\\ 
\rule{0pt}{3ex}
        & \bf{ (1,3,0)  (\textit{S}) }&  \boldmath{ $-2(x^2-5x+1+7x^3)/(x-1)^2$}  & \bf{0} & \bf{2} & \bf{0 } \\ 
\rule{0pt}{3ex}
           & $(3,3,-2/3)$+h.c. (\textit{U})& $-4x(-1+3x^3)/(x-1)^2$ &$\frac{24}{5}$ & 12 & 3\\ 
\rule{0pt}{3ex}
           & $(8,3,0)$ (\textit{Q}) & $-4(-x^2+2x-1+2x^3)/(x-1)^2$ &$0$ & 16 & 9\\ 
 \rule{0pt}{3ex}
           & $(1,2,3/2)$+h.c. (\textit{V})& $-4(-1+x+3x^2)/(x-1)$& $\frac{27}{5}$ & 1 & 0 \\  
\rule{0pt}{3ex}
           & $(6,2,-1/6)$+h.c. (\textit{Y}) & $4(-1+x+x^2)/(x-1)$ &$\frac{2}{5}$ & 6 & $10$\\ 
\rule{0pt}{3ex}
           & $(6,2,5/6)$+h.c. (\textit{B}) & $4(2x-1)/(x-1)$& $10$ & 6 & $10$\\           
\hline
\rule{0pt}{3ex}
$\Sigma, \overline{\Sigma}$ & $(1,3,-1)$ + h.c. (\textit{O}) & $-4 r_c x(4x^2-3x+1)/(x-1)^2$ & $\frac{18}{5}$ & 4  & 0\\
\rule{0pt}{3ex}
                                      & $(3,3,-1/3)$ + h.c. (\textit{P}) & $-2 r_c (7x^3-7x^2+5x-1)/(x-1)^2$ & $\frac{6}{5}$ & 12 & 3\\
\rule{0pt}{3ex}                                      
                                      & $(6,3,1/3)$ + h.c. (\textit{W}) & $-4 r_c (3x-1)(x^2-x+1)/(x-1)^2$ & $\frac{12}{5}$ & 24 & $15$\\
\rule{0pt}{3ex}
                                      & $(1,1,2)$ + h.c. (\textit{A}) & $-12 r_c x$ & $\frac{24}{5}$ & 0 & 0 \\
\rule{0pt}{3ex}
                                      & $(\overline{3},1,4/3)$ + h.c. (\textit{K}) & $-2 r_c (3x^2-6x+1)/(x-1)$ & $\frac{32}{5}$ & 0 & 1\\
\rule{0pt}{3ex}
                                      & $(\overline{6},1,2/3)$ + h.c. (\textit{M}) & $-4 r_c (1-3x)/(x-1)$ & $\frac{16}{5}$ & 0 & $5$\\
\rule{0pt}{3ex}
                                      & $(\overline{6},1,-1/3)$ + h.c. (\textit{L}) & $-2 r_c (x^2-7x+2)/(x-1)$ & $\frac{4}{5}$ & 0 & $5$\\
\rule{0pt}{3ex}
                                      & $(\overline{6},1,-4/3)$ + h.c. (\textit{N}) & $-4 r_c (x^2-4x+1)/(x-1)$ & $\frac{64}{5}$ & 0 & $5$\\
\rule{0pt}{3ex}
                                      & $(3,2,7/6)$ + h.c. (\textit{D}$_1$) & $-2 r_c (6x^3-10x^2+7x-1)/(x-1)^2$ & $\frac{49}{5}$ & 3 & 2\\
\rule{0pt}{3ex}
                                      & $(\overline{3},2,-1/6)$ + h.c. (\textit{E}$_1$) & $-2 r_c (4x^3-6x^2+5x-1)/(x-1)^2$ & $\frac{1}{5}$ & 3 & 2\\
\rule{0pt}{3ex}
                                      & $(\overline{3},2,-7/6)$ + h.c. (\textit{D}$_2$) & $-2 r_c (5x^3-8x^2+6x-1)/(x-1)^2$ & $\frac{49}{5}$& 3 & 2\\
\rule{0pt}{3ex}
                                      & $(8,2,1/2)$ + h.c. (\textit{C}$_1$) & $-2 r_c (3x^3-7x^2+8x-2)/(x-1)^2$ & $\frac{24}{5}$ & 8 & 12\\
\rule{0pt}{3ex}
                                      & $(8,2,-1/2)$ + h.c. (\textit{C}$_2$) & $-2 r_c (4x^3-9x^2+9x-2)/(x-1)^2$ & $\frac{24}{5}$ & 8 & 12\\                                     
\hline
\end{tabular}
\caption{Spectrum of unmixed states from the scalar representations. We have defined $r_c = 2\eta/\lambda$, 
and $\ell_i$ is the Dynkin index of each state ($\times2$ when there is a conjugate field) for the SM gauge group $i$, with GUT normalisation for $\ell_1$. The mass expressions correspond to the fermion masses,
and the scalar masses are obtained from Eq. (\ref{scalarmass}). The state which we will ultimately be most interested in, $S$, is highlighted in bold.}
\label{UnmixedScalars.TAB}
\end{table}

\begin{table}
\centering
\begin{tabular}{| c | c | c | c | c |}
\hline\hline
\textbf{Field} &  $(SU(3)_c,SU(2)_L,U(1)_Y)$  & $\ell_1$ & $\ell_2$ & $\ell_3$\\
\hline 
\hline
\rule{0pt}{3ex}
$H$, $\Phi$, $\Sigma\overline{\Sigma}$ & $(1,2,1/2)$+h.c.  (\textit{h}$_{1-4}$)& $\frac{3}{10}$ & $\frac{1}{2}$ & 0\\
\hline
\rule{0pt}{3ex}
$H$, $\Phi$, $\Sigma\overline{\Sigma}$ & $(3,1,-1/3)$+h.c.  (\textit{T}$_{1-5}$)& $\frac{1}{5}$ & 0 &$\frac{1}{2}$\\
\hline
\rule{0pt}{3ex}
$\Phi$, $\Sigma\overline{\Sigma}$ & $(3,1,2/3)$+h.c.  ($J_{1-3}$)& $\frac{4}{5}$ & 0 &$\frac{1}{2}$\\
\hline
\rule{0pt}{3ex}
$\Phi$, $\Sigma\overline{\Sigma}$ & $(1,1,+1)$+h.c.  ($F_{1-2}$)& $\frac{3}{5}$ & 0 & 0 \\
\hline
\rule{0pt}{3ex}
$\Phi$ & $(8,1,0)$ ($R_{1,2}$)&  $0$ & 0 & 3\\
\hline
\rule{0pt}{3ex}
$\Phi$, $\Sigma\overline{\Sigma}$ & $(3,2,+1/6)$+h.c.  ($E_{2-4}$)& $\frac{1}{10}$ & $\frac{3}{2}$ & 1\\
\hline
\rule{0pt}{3ex}
$\Phi$, $\Sigma\overline{\Sigma}$ & $(3,2,-5/6)$+h.c.  ($X_{1-2}$)& $\frac{5}{2}$ & $\frac{3}{2}$ & 1\\
\hline
\rule{0pt}{3ex}
$\Phi$, $\Sigma\overline{\Sigma}$ & $(1,1,0)$  ($G_{1-5}$)& $0$ & 0 & 0\\
\hline
\end{tabular}
\caption{Mixed states from the scalar representations and their Dynkin indices $\ell_i$, with GUT normalisation for $\ell_1$. There are no compact expressions for the masses of the mixed states.}
\label{MixedScalars.TAB}
\end{table}

The spontaneous symmetry breaking of SO(10) to the SM consumes 33 massless Nambu-Goldstone bosons, yielding
33 massive gauge bosons as calculated in \cite{AG}, which we reproduce in Table \ref{GaugeBosons.TAB}. 
In the Table and elsewhere, $g_U$ refers to the SO(10) unified gauge coupling.

\begin{table}
\centering
\resizebox{\columnwidth}{!}{
\begin{tabular}{| c | c | c | c | c | c |}
\hline\hline
\textbf{Field} &  $(SU(3)_c,SU(2)_L,U(1)_Y)$ & \textbf{Mass /}$\mu_\Phi$ & $\ell_1$ & $\ell_2$ & $\ell_3$\\
\hline 
\hline
$W_R^0$ & $(1,1,0)$ & $\sqrt{10}~g_U\left(\frac{2}{\eta \lambda}\frac{x(1-3x)(1+x^2)}{(1-x)^2} \right)^{1/2}$ & 0 & 0 & 0\\
\hline
$X_{PS}$ & $(3,1,2/3) +$ h.c. &$g_U \left(4 \left|\frac{(1-3 x) x (x^2+1)}{(1-x)^2 \eta  \lambda}\right| +8 \left|\frac{-x^2-2 x+1}{(1-x) \lambda }\right|^2+16  \left|\frac{x}{\lambda }\right|^2 \right)^{1/2}$ & $\frac{4}{5}$ & 0 & $\frac{1}{2}$\\
\hline
$W_R^{\pm}$ & $(1,1,+1) +$ h.c. &$g_U \left(4 \left|\frac{(1-3 x) x (x^2+1)}{(1-x)^2 \eta  \lambda}\right|+24  \left|\frac{x}{\lambda }\right|^2 \right)^{1/2}$ & $\frac{3}{5}$ & 0 & 0\\
\hline
$X',~Y'$ & $(3,2,1/6) +$ h.c. &$g_U \left(4 \left| \frac{(1-3 x) x \left(x^2+1\right)}{(1-x)^2 \eta  \lambda }\right| +4 \left| \frac{x}{\lambda }-\frac{-x^2-2 x+1}{(1-x) \lambda }\right|^2+2 \left| \frac{x \left(1-5 x^2\right)}{(1-x)^2 \lambda }-\frac{x}{\lambda }\right|^2 \right)^{1/2}$ & $\frac{1}{10}$ & $\frac{3}{2}$ & 1\\
\hline
$X,~Y$ & $(3,2,-5/6) +$ h.c. &$g_U \left(4 \left| -\frac{x}{\lambda }-\frac{-x^2-2 x+1}{(1-x) \lambda }\right|^2+2 \left| -\frac{x}{\lambda }-\frac{\left(1-5 x^2\right) x}{(1-x)^2 \lambda }\right|^2\right)^{1/2}$ & $\frac{5}{2}$ & $\frac{3}{2}$ & 1\\
\hline
\end{tabular}
}
\caption{The mass spectrum of GUT gauge bosons, and their Dynkin indices under each SM 
gauge group, $\ell_i$.}
\label{GaugeBosons.TAB}
\end{table}

An important difference between the present analysis and the previous works in \cite{bmsv,AG} is the scale of supersymmetry breaking.
In previous works, the supersymmetry breaking scale, ${\widetilde m}$ was assumed to be very small compared with the mass scales associated with
the {\bf 126} and {\bf 210} Higgses. In contrast, here we are 
interested in the case where the supersymmetry-breaking scale is very high, to the point that it can be as large as $\widetilde{m}\sim 0.1 \mu_\Phi$.
As a result, we must account for non-negligible supersymmetry breaking corrections to the GUT sparticle spectrum. 
This will alter the results for the spectra considered above \cite{bmsv,AG} as we now discuss.

The unmixed fermion masses (listed in Table \ref{UnmixedScalars.TAB}) remain the same as in \cite{bmsv,AG}, 
since these are the Higgsinos associated with the GUT scalar Higgses, and therefore do not receive corrections from SUSY breaking.
Rather, because the ``$\mu$-term" parameters, $\mu_\Phi$, and $\mu_\Sigma$ are GUT scale, they generically lead to  large Higgsino masses. 
If we could ignore supersymmetry breaking, then tuning $x$ so that a particular unmixed state becomes light would
result in both a light scalar and a light fermion. However, when supersymmetry breaking is comparable (or at least non-negligible) to $\mu_\Phi$,
the tuning of $x$ resulting in a light scalar is altered and the fermion partner will in general remain heavy. To see this, recall that
unmixed scalar masses receive corrections of the following form:
\beq
m_{S_i} = \left(m_{F_i}^2(x) + \widetilde{m}^2 \right)^{1/2}~,
\label{scalarmass}
\eeq
so that now minimizing $x$ to set a particular scalar mass to zero does not simultaneously set the accompanying fermion mass to zero.

Determining the masses of the mixed states is somewhat more complicated. 
Mixed fermion masses are no longer obtained by simply diagonalising the matrices given in \cite{AG}, since there are soft SUSY mass terms for the gauginos. 
Thus, the diagonal entry of the matrices in \cite{AG} corresponding to $m_{\tilde{G}\tilde{G}}$ receives a correction of size $m_{1/2} \sim \widetilde{m}$. 
For example, let us consider the matrix for the fermion states, with charges $(1,1,-1)$ under ${\rm SU(3)}_c$, ${\rm SU(2)}_L$ and ${\rm U(1)}_Y$ which becomes:
\begin{align}
\mathcal{F}^F = \bmat 2( \mu_\Sigma + \eta(v_{1,1,1}+3 v_{1,1,15} )) & -2i \sqrt{3} \eta \sigma_{1,3,10} &  - g_U \sqrt{2} \sigma_{1,3,\overline{10}}^*
\\
2i \sqrt{3} \eta \sigma_{1,3,\overline{10}} & 2(\mu_\Phi + \lambda(v_{1,1,1}+2 v_{1,1,15})) & \sqrt{24} i g_U v_{1,3,15}^* 
\\
- g_U \sqrt{2} \sigma_{1,3, 10}^* & -\sqrt{24} i g_U v_{1,3,15}^* & m_{1/2}
\emat~,
\label{FFmatrix}
\end{align}
in the ($\Sigma \overline{\Sigma} (1,3,10), \Phi (3,1,15), {\tilde W_R^\pm}$) basis. 
Mixed scalar masses are also shifted, but now they are obtained by diagonalizing $(M^\dagger M)_{n \times n} + \widetilde{m}^2 \mathbb{1}_{n-k \times n-k}$, where $k$ is the number of Nambu-Goldstone states, while ensuring that the Nambu-Goldstone bosons remain massless. Once again, tuning a scalar to be light 
leaves its fermionic partner heavy (of order $\mu_\Phi$ or $m_{1/2}$). We assume universal soft SUSY-breaking masses for the scalars. The masses of the gauge bosons from the GUT spectrum do not change in the presence of SUSY breaking. In the following, with the exception of the $G$ state, we use the mixed states (and their masses) only in the threshold corrections necessary for obtaining gauge coupling unification, and will not be considered as candidates for a possible light scalar.
We will, however, examine the case that one of the five $G$ states listed in Table~\ref{MixedScalars.TAB} is left light.

As we have noted above, the massless spectrum is obtained by varying $x$ such that one or more states are massless. 
Numerically, we proceed by solving for $x_0$, the value of $x$ for which a particular unmixed scalar state is massless. 
Without taking SUSY breaking effects into account, there might have been multiple states for which the same $x_0$ solved $m_i(x_0)=0$. 
However, after accounting for SUSY breaking, there is only ever one scalar which is massless for a particular choice of $x_0$. Two distinct sets of solutions for massless states are found, depending on whether we take the SUSY mass-squared contribution to the SO(10) scalars to be positive or negative. 
We will consider both possibilities in our analysis. Negative SUSY mass-squared contributions to the SO(10) scalars do not lead to tachyons as long as the GUT-scale mass is larger, which it invariably is.

Additionally, one might have worried that a particular choice of $x_0$ would simultaneously set a GUT gauge boson mass to zero, which could be dangerous for proton decay. The values of $x_0$ for which a proton-decay-mediating GUT gauge boson is massless are
\beq
x_0^{\times} = \left\{-1,~\frac{1}{3},~\frac{1}{2}  \right\}\ .
\eeq
However, since the SUSY breaking shifts the scalar masses but not the GUT gauge bosons, this never occurs, and the perturbation away from a value $x_0^{\times}$ due to SUSY breaking is sufficiently large that the GUT gauge bosons have masses of $\mathcal{O} (\mu_\Phi)$.

Note that a second tuning is also required in this model to obtain a light Higgs scalar doublet
(equivalent to the SU(5) doublet-triplet splitting). In the supersymmetric limit the Higgs doublets 
$H_{u,d}$ are identified as a linear combination of the SU(2)$_L$ doublets appearing in the representations (\ref{10rep})-(\ref{210rep}), by requiring that the corresponding mass matrix in the scalar potential has zero determinant. Using the superpotential (\ref{eq:superpot}), this leads to a condition~\cite{AG,bmsv}
\beq
     \mu_H = \mu_\Phi \frac{\alpha {\overline\alpha}}{2\eta\lambda} \frac{p_{10}}{(x-1)p_3p_5}~,
     \label{eq:muH}
\eeq
where $p_{3,5,10}$ are polynomials of $x$ defined in Appendix C of \cite{bmsv}. This condition will be modified when we consider supersymmetry breaking, since the scale of supersymmetry breaking in our model is large ${\widetilde m}\simeq 0.1 \mu_\Phi \simeq 10^{14}$ GeV.
In principle, we must tune the $4\times4$ mass-squared matrix for the Higgs doublets in $H, \Sigma, \overline{\Sigma}$, and $\Phi$ with $Y = 1/2$ 
corresponding to state ($h$) in Table \ref{MixedScalars.TAB}. 
Note that only the doublet in the $(2,2, \overline{10})$ mixes with the doublet with $Y = 1/2$ in $H$ as the other doublet in the $(2,2,10)$ has hypercharge $Y = 3/2$. 
In the following numerical analysis, we impose the zero-determinant condition for ${\cal B}^{h\dagger}{\cal B}^h$ with the inclusion of the supersymmetry breaking mass, $\widetilde m$, namely, ${\rm det}({\cal B}^{h\dagger}{\cal B}^h
-|\widetilde m|^2\mathbb{1})=0$, where ${\cal B}^h$ is the $4\times4$ mass matrix for the Higgs boson doublets in $H, \Sigma, \overline{\Sigma}$ and $\Phi$ states, given by
\begin{eqnarray}
{\cal B}^h = 
\left(
   {\scriptsize
   \begin{array}{cccc}
   -\mu_H & \bar\alpha\sqrt{3}(v_{1,3,15}-v_{1,1,15}) & -\alpha\sqrt{3}(v_{1,3,15}+v_{1,1,15}) & -\bar\alpha\sigma_{1,3,\overline{10}} \\
   -\bar\alpha\sqrt{3}(v_{1,3,15}+v_{1,1,15}) & 0 & -(2\mu_\Sigma+4\eta(v_{1,1,15}+v_{1,3,15}) )& 0 \\
   \alpha\sqrt{3}(v_{1,3,15}-v_{1,1,15}) &  -(2\mu_\Sigma+4\eta(v_{1,1,15}-v_{1,3,15}) )& 0 & -2\eta\sigma_{1,3,\overline{10}}\\
   -\alpha\sigma_{1,3,10} & -2\eta\sigma_{1,3,10} & 0 & -2\mu_{\Phi}+6\lambda(v_{1,3,15}-v_{1,1,15})
   \end{array}
   }
\right),
\label{eq:Bh}
\end{eqnarray}
in the $(H(2,2,1), \overline{\Sigma}(2,2,15),\Sigma(2,2,15),\Phi(2,2,\overline{10}))$ basis.
In practice, at scales below $\widetilde m$, we include the SM particles and one light GUT scalar representation in the renormalization group evolution of the gauge coupling. The lightest Higgsino starts to contribute at the scale 
$\widetilde m$ where all the other MSSM particles and the fermion partners of the lightest scalar state must be included in the running.

\section{Running of gauge couplings}

One of the motivations for low energy supersymmetry is the unification of the gauge couplings at high energy \cite{Ellis:1990zq}. In the absence of large threshold
corrections, running up the gauge couplings in the SM does not lead to unification, or alternatively, running down a unified gauge coupling in the SM leads to low energy gauge couplings which do not all agree with experiment.
Additional states in a supersymmetric theory, which are not in complete SU(5) multiplets, alter the running in such a way that allows for gauge coupling unification. Of course as the scale of supersymmetry breaking approaches the GUT scale, we recover the SM limit and lose the unification prediction. In a large GUT such as SO(10),
the size of the representations needed to break SO(10) down to the SM would indicate that threshold corrections can not be ignored.
For a given value of $x$ (and couplings $\eta, \lambda$ and mass parameters $\mu_\Phi, \mu_\Sigma$, though they are related through the
last expressions in Eq.(\ref{sol})), the superheavy spectrum is known and the threshold corrections can be computed. 
As we show below, these are sufficiently large that unification can be achieved even in high-scale supersymmetry models. 
However, if all states beyond the SM are superheavy, there is no possibility to prevent the Higgs quartic coupling from running negative.

Instead, Higgs vacuum stability and gauge coupling unification may be achieved if a state in one of the Higgs representations remains relatively light. If the light GUT state is not a SM singlet, we expect two changes in the gauge coupling running. First the running proceeds via the SM $\beta$-functions up to the mass threshold of the light GUT state. Above this threshold we then have modified $\beta$-functions from the extra degrees of freedom of the light GUT state which are run up to the supersymmetry breaking scale. Above this scale, the renormalization group equations (RGEs) are further modified and the now twice modified RGEs are then run up to the GUT scale, which can be defined once GUT threshold corrections are included. 

The running of the SM couplings is well known up to three loops, and the $\beta$-functions can be found in for example \cite{SMRGE, Buttazzo:2013uya}. At one loop, the SM gauge coupling $\beta$-function coefficients are:
\beq
b_1^{SM} = \frac{41}{10}, ~~~b_2^{SM} = -\frac{19}{6},~~~b_3^{SM} = -7 \ .
\eeq
At the mass threshold of the light GUT state, the running of the gauge couplings is altered, such that at one loop, the $\beta$-function coefficients are: 
\beq
b_1^{LS} = \frac{41}{10} + \frac{1}{3}\ell_1^i, ~~~b_2^{LS} = -\frac{19}{6}+ \frac{1}{3}\ell_2^i,~~~b_3^{LS} = -7 + \frac{1}{3}\ell_3^i \ ,
\label{LSrunning.EQ}
\eeq
where $\ell^i_a$ is the Dynkin index under the gauge group $a$ of the $i$-th GUT state, and the factor of $1/3$ is because the GUT state is a complex scalar. 

Once one crosses the threshold of the SUSY states, the running is again changed. The first change is due to the necessary switch from the $\MS$ regularisation scheme to the $\DR$ scheme, since the former does not preserve SUSY. Additionally, one notes that the fermionic superpartner of the light GUT state will have a mass of order $\widetilde m$, and must therefore also be taken into account\footnote{Typically, we assume ${\widetilde m} \ge m_I = 3 \times 10^{13}$ GeV.}. This will lead to an additional factor of $\frac{2}{3}\ell^i_a$, since the GUT state superpartner is a Weyl fermion. At one loop, the $\beta$-function coefficients are now:
\beq
b_1^{SUSY} = \frac{33}{5} + \ell_1^i, ~~~b_2^{SUSY} = 1+ \ell_2^i,~~~b_3^{SUSY} = -3 + \ell_3^i \ ,
\label{SUSYrunning.EQ}
\eeq
where we have included the superpartner contribution to all SM states. 
Note that in the following analysis the MSSM superpartners are assumed to be degenerate in mass so that their threshold corrections at $\widetilde m$ vanish.
The Dynkin indices of each of the scalar GUT states are listed in Tables \ref{UnmixedScalars.TAB} and \ref{MixedScalars.TAB}, so that the modifications to the running of the gauge couplings for each of the solutions in the previous section can be implemented with ease.

\section{GUT-scale threshold corrections}

As noted above, once the supersymmetry breaking threshold is passed, running to the GUT scale
proceeds including the MSSM and the light scalar determined by a given value of $x_0$. The precise value of the GUT scale
will depend on additional threshold corrections from the remaining states in the {\bf 126} and {\bf 210} Higgses 
(and Higgsinos) and GUT scale gauge bosons (and gauginos).
More specifically, high-scale states for a given value of $x_0$ will tend to have non-degenerate masses, and as such will result in non-zero threshold corrections at the GUT scale which must be taken into account. The threshold corrections $\lambda_i$ are defined in terms of the gauge coupling at a given scale $\mu_*$ (taken to be $\mu_\Phi$), and the putative unified coupling at such a scale, in the appropriate regularisation scheme:
\beq
\left(\frac{1}{g_i^2(\mu_*)}\right)_{\MS, \DR} = \left(\frac{1}{g_U^2(\mu_*)} \right)_{\MS, \DR} - \left(\frac{\lambda_i}{48\pi^2}\right)_{\MS, \DR} \ ,
\eeq
where one can then calculate $\lambda_i$ in terms of the masses of the heavy states, and is found to be \cite{Hall:1980kf, Langacker:1992rq}
\begin{align}
\left(\lambda_i\right)_{\MS} &= \ell_i^V - 21 \ell_i^V \ln \frac{M_V}{\mu} + c_S \ell_i^S \ln \frac{M_S}{\mu} + c_F \ell_i^F \ln \frac{M_{F}}{\mu} \ , \\
\left(\lambda_i\right)_{\DR} &= - 21 \ell_i^V \ln \frac{M_V}{\mu} + c_S \ell_i^S \ln \frac{M_S}{\mu} + c_F \ell_i^F \ln \frac{M_{F}}{\mu} \ ,
\label{DRthresholds.EQ}
\end{align}
where $\ell_i^x$ are the Dynkin indices corresponding to ultra-heavy massive vector bosons ($V$), scalars ($S$) or fermions ($F$). The coefficient $c_S = 1,~2$ for real and complex scalars respectively, while $c_F = 4,~4,~8$ for Weyl, Majorana and Dirac fermions respectively. 

Unfortunately, given that $g_U$ is a deep UV quantity, and we live in the IR, we cannot unambiguously define $g_U$ from our perspective. Any number of definitions can be proposed, such as choosing $g_U(M_U) = g_2(M_U)= g_1(M_U)$, or $g_U(M_U) = g_2(M_U)= g_3(M_U)$, but none of these are necessarily correct. Instead $g_U$ can only be correctly defined from the UV perspective in the GUT phase of the theory. Then, at a given scale $M_*$, one can match to the broken-GUT phase with the couplings $g_1$, $g_2$ and $g_3$. This matching will likely involve substantial changes from the threshold corrections, such that any $g_i(M_*)$ may be quite different from $g_U(M_*)$. Therefore, for an analysis of how unification is achieved as calculated in the IR, without knowledge of a specific UV completion, one would prefer to abstain from defining the unified coupling. 

We may use the prescription proposed in \cite{Ellis:2015jwa}, which allows one to assess the quality of gauge coupling unification in the presence of threshold corrections, without substantial impact from the definition of $g_U$. This prescription calls for the definition of quantities which are independent of the unified gauge coupling at a scale $\mu$, $g_U(\mu)$. We define these quantities as
\begin{align}
\left( \frac{\Delta \lambda_{ij}(\mu)}{48\pi^2}\right)_{\MS,\DR} \equiv \left( \frac{1}{g_i^2(\mu)}-\frac{1}{g_j^2(\mu)}\right)_{\MS,\DR} = \left( \frac{\lambda_j(\mu) - \lambda_i(\mu)}{48\pi^2}\right)_{\MS,\DR} \ ,
\label{lij}
\end{align}
such that only two need be defined so as to specify the GUT matching conditions. Thus, in the IR we may calculate the required $\Delta\lambda_{ij}$ at any scale, and compare with the $\Delta\lambda_{ij}$ which are obtained in the UV for a specific GUT spectrum. If the $\Delta\lambda_{ij}$ in the IR and the UV match, then it is possible that unification is achieved. The differences in the required threshold corrections, as viewed from the IR, contain an ambiguity since they may not account for a constant term which cancels. Therefore matching the IR and UV calculations of $\Delta\lambda_{ij}$ specifies
\begin{align}
\frac{1}{g_U^2(\mu)} + C \quad \text{and} \quad \frac{\lambda_i(\mu)}{48\pi^2} + C \ ,
\end{align}
where $C$ is a constant shift. Since both of these quantities are a priori known from the UV perspective, specifying the UV theory allows for the ambiguity to be resolved. 

Our prescription for finding solutions which lead to potentially acceptable gauge coupling unification is outlined as follows. 
We start with the SM at low energies, supplemented with a light GUT state corresponding to one of the possibilities listed in 
Tables \ref{UnmixedScalars.TAB} or \ref{MixedScalars.TAB}. At each renormalization scale $\mu$, we can calculate the quantities
$\Delta \lambda_{ij}$ using the left hand side of Eq.\,(\ref{lij}). We will assume that the supersymmetric particle spectrum lies at $3 \times 10^{13}$ GeV, and above that scale
the $\Delta \lambda_{ij}$ are computed in the MSSM plus the additional light scalar (and fermion superpartner). 
Next, we scan over the couplings $\lambda, \eta$. Recall that $x$ is fixed by requiring that one of the scalars is light using Eq.\,(\ref{scalarmass})
with $m_F$ a function of $x$ taken from Table \ref{UnmixedScalars.TAB}. We are then left with three unknowns:
$\mu_\Phi, {\widetilde m}$, and $g_U$, all of which are needed to determine the masses of the heavy states participating in the threshold corrections\footnote{Once $x$ is determined to obtain a light state, the remaining superpotential parameter $\mu_\Sigma$ is fixed by Eq. (\ref{sol}) when $\lambda,~\eta$, and $\mu_\Phi$ are input.}.
For given values of these three parameters, the threshold corrections in Eq. (\ref{DRthresholds.EQ}) can be computed, as can their differences
given in the right-hand side of Eq.~(\ref{lij}). Comparing these two results for $\Delta \lambda_{ij}$, we can determine the degree to which a solution
is acceptable.  In other words, viability is determined by scanning $\Delta\lambda_{ij}$ in the allowed $\eta-\lambda$ parameter space for each of the light state solutions in section \ref{GUTfields.SEC} above, and comparing with the required $\Delta\lambda_{ij}$ calculated in the IR from the running of the gauge couplings towards the UV.

To find viable unification solutions, we search a set of parameters, $(g_U,~\mu_\Phi,~\widetilde m,~m_\chi)$, for a given $(\lambda,~\eta)$ so that the function $\chi^2$ defined by
\beq
\chi^2(g_U,\mu_\Phi,\widetilde m,m_\chi) \equiv
\sum_{i=1}^3 \left[ g_i^{-2}(\mu_\Phi)  - \left(g_U^{-2} - \frac{\lambda_i(g_U,\mu_\Phi,\widetilde m,m_\chi)}{48\pi^2}\right) \right]^2/\sigma_{i}^2,
\label{chi2}
\eeq
is minimized, where $m_\chi$ is the mass of the light scalar state, and $\sigma_i^2\equiv \sigma^2_{g_i^{-2}}+\sigma_{{\rm th},i}^2$ with $\sigma^2_{g_i^{-2}}$ and $\sigma_{{\rm th},i}$ being the experimental errors for $g_i^{-2}(m_Z)$ and theoretical uncertainties, respectively. For the theoretical uncertainties, arising from the matching scale dependence of $g_i$ and $\lambda_i$,
we assume 1\% of  $ [g_i^{-2}+\lambda_i/(48\pi^2)] \equiv \hat{\lambda}_i$, as an estimate of the next-order corrections to the couplings and thresholds.
We summarize in Table~\ref{tab:parameters},  taken from \cite{Patrignani:2016xqp}, the values of the input parameters we have used for the tree level couplings given by
\beq
g_1 = \sqrt{\frac{5}{3}}g_Y,~g_Y = 2(\sqrt{2}G_F)^{1/2}\sqrt{m_Z^2-m_W^2},~g_2 = 2(\sqrt{2}G_F)^{1/2}m_W,~y_t = 2(G_F m_t^2/\sqrt{2})^{1/2}.
\eeq
In our analysis, we have used the two-loop corrected couplings at $\mu=m_t$ \cite{Buttazzo:2013uya}, 
\beq
g_1(m_t) = 0.4626,~g_2(m_t) = 0.6478,~g_3(m_t) = 1.166,~y_t(m_t) = 0.9379,
\eeq
with uncertainties $\sigma_{g_1^{-2}}=2.434\times10^{-3},~\sigma_{g_2^{-2}}=1.191\times10^{-3},~\sigma_{g_3^{-2}}=7.437\times10^{-3}$, 
which are however minor contributions in $\sigma_i$ compared to the 1\% theoretical uncertainties.
Note that our definition of $\chi^2$ represents an underdetermined system and we might expect that there are multiple (or continuous families of) solutions giving $\chi^2 = 0$. 

\begin{table}[tb]
\centering
\begin{tabular}{lll}
\hline
\hline
$W$ boson mass & $m_W$ & $80.385(15)$ GeV\\
$Z$ boson mass & $m_Z$ & $91.1876(21)$ GeV\\
Higgs boson mass & $m_h$ & $125.18(16)$ GeV\\
top quark pole mass & $m_t$ & $173.5(1.1)$ GeV\\
Fermi constant & $G_F$ & $1.166387(6)\times10^{-5}$ GeV$^{-2}$\\
strong coupling constant & $\alpha_s(m_Z)$ & $0.1182(16)$
\\\hline
\hline
\end{tabular}
\caption{The physical constants we have used for the input parameters are summarized.}
\label{tab:parameters}
\end{table}

As noted above, the values of the gauge couplings are also affected by $\widetilde m$ and $m_\chi$ through the changes of the $\beta$-functions.
The threshold corrections $\lambda_i$ evaluated at $\mu=\mu_\Phi$ are now the functions of $g_U,~\mu_\Phi,~\lambda,~\eta$ and $\widetilde m$, and include all the contributions from the GUT scale Higgs, Higgsino and gaugino, except for one Higgs doublet and the lightest GUT scalar state.
It is worth noting that any parameter sets that make $\chi^2$ close to zero also satisfies the second equality in (\ref{lij}).

\section{Unification Solutions}

In this section, we discuss the few solutions which lead to acceptable gauge coupling unification, as described above.
The numerical procedure for the analysis is as follows: we run the gauge, top and Higgs quartic couplings up in the SM at the 2-loop level in $\MS$. At the scale of the light GUT state, we match to the new running in $\MS$. 
In principle this matching should be done including both $\log$ and finite 1-loop threshold corrections. However, 
for all the solutions for the unmixed states listed above, there is only one light state, so that any $\log$ effects are zero, since we can match at the scale of the new state exactly. We have not implemented any possible finite threshold corrections.  We then run up from the light state scale in $\MS$ at the 2-loop level, using the modified 1-loop $\beta$-function coefficients as defined in Eq.\,(\ref{LSrunning.EQ}). 
We then match to the SUSY scale, $\widetilde{m}$, assuming all SUSY states are degenerate in mass, so again there are no logarithmic threshold corrections. 
We match ${g_i}_{\MS}$ to ${g_i}_{\DR}$ at this scale, and perform the calculations of the running and threshold corrections above this scale in $\DR$ to preserve supersymmetry. 
We use the usual MSSM 2-loop RGEs, with the modified 1-loop $\beta$-functions as in Eq. (\ref{SUSYrunning.EQ}). 
The GUT scale thresholds are computed using Eq.\,(\ref{DRthresholds.EQ}).

Among 21 unmixed scalar states, only the $(1,3,0)$ state in $\Phi$ labeled by $S$, and shown in boldface in Table 
\ref{UnmixedScalars.TAB}, can serve as a promising light state to achieve phenomenologically viable unification.
For this state to have mass as low as ${\cal O}(0.1-1)$ TeV,  the solution with $x\simeq 0.63$ in 
the $\widetilde m=0$ limit leads to a viable parameter space.
We have also examined the case that one of the five $G$ boson states listed in Table \ref{MixedScalars.TAB} is light as well as the case with no extra light states, even though the latter can not help resolve the problem of 
the Higgs quartic coupling and vacuum stability. 
Under the restriction that ${\widetilde m} > 3 \times 10^{13}$ GeV, other than $S$,  none of the other light state solutions leads to unification within 3$\sigma$ 
(where $\sigma$ is determined from Eq. (\ref{chi2})).

\begin{figure}[ht!]
\centering
\includegraphics[scale=0.75]{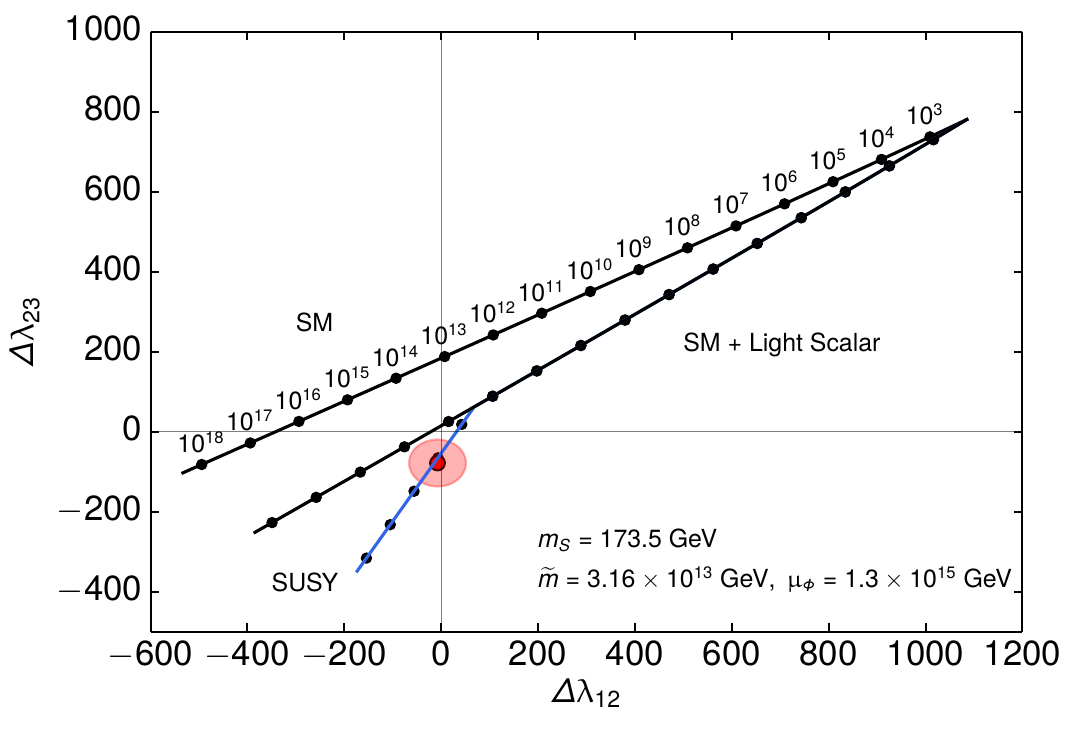}
\caption{$\Delta\lambda$ plot for $S$ ($\Phi(1,3,0)$) as the only light state. We have set $x_0 \sim 0.63$. The lowest branch shows the evolution after matching to SUSY at $10^{13.5}$ GeV.
The red circle corresponds to the nearly best fit point (with acceptable proton lifetime).}
\label{130_Delta_Lambda_10to4.FIG}
\end{figure}

\begin{figure}[ht!]
\centering
\includegraphics[scale=0.6]{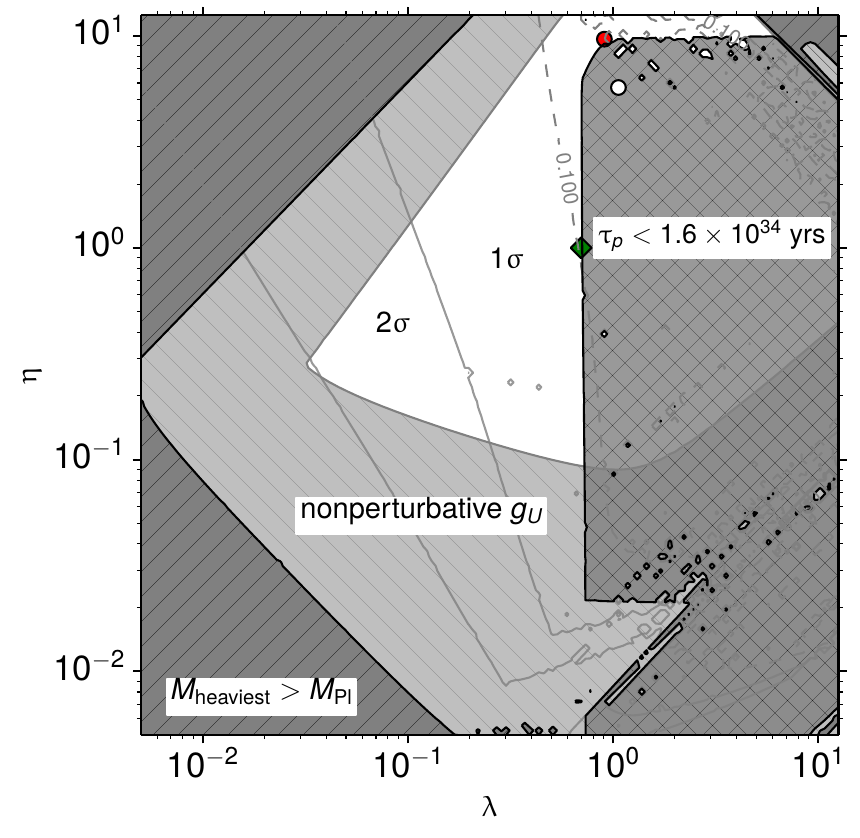}
\caption{A plot of the $(\lambda,\eta)$ plane showing the allowed region (in white) for gauge coupling unification with a light $S$ state. 
The light gray region is not allowed because $g_U$ becomes nonperturbative below the mass scale of the heaviest state. The dark gray regions hatched by diagonal lines are disfavored due to the presence of the heaviest GUT state being greater than $M_{\rm Pl}$. The dark gray region hatched by crossed lines is excluded by limits on proton decay. The white (red) circle indicates the best fit (viable) points, and the green diamond corresponds to the sample point given in Table~\ref{points.TAB} exhibiting smaller values of $\lambda$ and $\eta$.}
\label{chi2.FIG}
\end{figure}

In Figure \ref{130_Delta_Lambda_10to4.FIG}, 
we show the values of $\Delta \lambda_{12}$ and $\Delta \lambda_{23}$ parametrically
as a function of the renormalization scale corresponding to the state $S$.
The upper line  shows the evolution assuming only SM content. 
The left-hand side of Eq. (\ref{lij}) is used to calculate
$\Delta \lambda_{ij}$ with the SM running of the gauge couplings $g^2_i$ and $g^2_j$. Branching from the SM line is a line
appearing steeper in the ($\Delta \lambda_{12}$, $\Delta \lambda_{23}$) plane, which is computed assuming the SM plus the single light scalar
for which we have assumed $m_\chi = m_S = m_t$, though the results are very insensitive to the exact value of $m_S$. 
This line is deflected at $\mu = {|\widetilde m}|$ due to the appearance of SUSY states, so that above this scale the $\Delta \lambda_{ij}$ are computed
using the full MSSM spectrum plus our light scalar (shown as a blue line). 
The black-filled circles tracking the lines show the value of the renormalization scale $\mu$ in units of GeV, which varies from $10^3$ GeV to $10^{18}$ GeV.
The larger red-filled circle, surrounded by a pink shaded region,
corresponds to a point very near the best fit found by minimizing the $\chi^2$ function in Eq. (\ref{chi2}) and shown in the ($\lambda, \eta$) plane in Figure  \ref{chi2.FIG}. 
The values for the threshold corrections for this point are: $\lambda_ 1 = 2741.3, \lambda_2 = 2733.7$ and $\lambda_3 = 2655.0$ and thus
this point sits at $\Delta \lambda_{12} = -7.36 \pm 58.9$ and $\Delta \lambda_{23} = -78.6 \pm 58.9$, with the errors determined from $0.01 (48 \pi^2) (\hat{\lambda}_i^2 
+ \hat{\lambda}_j^2)^{1/2}$ and are reflected in the size of the pink shaded region.
In Figure  \ref{chi2.FIG}, the white circle corresponds to the actual best  fit but has a proton lifetime which is 
slightly below the experimental limit (see below for more detail). 
The fact that this point lies on the parametric line indicates a near perfect fit (with $\chi^2 = 0$, as we anticipated from an underdetermined system).
The parameters associated with the red point are given in Table \ref{points.TAB}.
Since the value of $\eta$ at the best fit point is large and close to the nonperturbative limit, we also show in Figure~\ref{chi2.FIG} a sample point with a smaller value of $\eta$, corresponding to the value listed in Table~\ref{points.TAB}.
As seen in Figure~\ref{chi2.FIG} and discussed in section~\ref{Best fit point}, points with smaller, more perturbative values of 
$\eta$, also exhibit satisfactory gauge coupling unification, and the value of $\eta$ turns out to be irrelevant for the issue of the Higgs stability and radiative breaking of electroweak symmetry, as will be discussed in section~\ref{Electroweak vacuum stability}.

\begin{table}[ht]
\centering
\resizebox{\columnwidth}{!}{
\begin{tabular}{| c | c | c | c | c | c | c | c |}
\hline\hline
& $(\lambda,\eta)$ &  $\chi^2$  & $m_\chi/{\rm GeV}$ & $\widetilde m/10^{13}~{\rm GeV}$ & $\mu_\Phi/10^{15}~{\rm GeV}$ & $g_U(\mu_\Phi)$ & $\tau_p/10^{34}~{\rm years}$ \\
\hline 
\hline
\rule{0pt}{3ex}
best fit ($S$)& \multirow{2}{*}{$(0.9082, 9.663)$} & \multirow{2}{*}{$8.2\times10^{-3}$} & \multirow{2}{*}{173.5} & \multirow{2}{*}{3.16} & \multirow{2}{*}{1.3} & \multirow{2}{*}{0.3373} & \multirow{2}{*}{1.8}\\
(w/ $\tau_p$ limit) & & & & & & &\\
best fit ($S$)& \multirow{2}{*}{$(1.063, 5.713)$} & \multirow{2}{*}{0} & \multirow{2}{*}{173.5} & \multirow{2}{*}{3.17} & \multirow{2}{*}{1.3} & \multirow{2}{*}{0.3489} & \multirow{2}{*}{0.84}\\
(w/o $\tau_p$ limit) & & & & & & &\\ 
sample point ($S$)& {$(0.7, 1.0)$} & {0.35} & {173.5} & {3.16} & {1.3} & {0.3789} & {4.1}\\ \hline
best fit ($E$)& \multirow{2}{*}{$(1.245, 9.663)$} & \multirow{2}{*}{11.37} & \multirow{2}{*}{$10^{10}$} & \multirow{2}{*}{3.16} & \multirow{2}{*}{1.0} & \multirow{2}{*}{0.3697} & \multirow{2}{*}{1.8}\\
(w/ $\tau_p$ limit) & & & & & & &\\
best fit ($E$)& \multirow{2}{*}{$(9.169,9.663)$} & \multirow{2}{*}{7.035} & \multirow{2}{*}{$3.2\times10^{11}$} & \multirow{2}{*}{3.16} & \multirow{2}{*}{1.0} & \multirow{2}{*}{0.4104} & \multirow{2}{*}{$6.5\times10^{-4}$}\\
(w/o $\tau_p$ limit) & & & & & & &\\
\hline
best fit (SM)& \multirow{2}{*}{$(0.1170, 3.753)$} & \multirow{2}{*}{12.46} & \multirow{2}{*}{-} & \multirow{2}{*}{3.16} & \multirow{2}{*}{1.0} & \multirow{2}{*}{0.3324} & \multirow{2}{*}{2.05}\\
(w/ $\tau_p$ limit) & & & & & & &\\
best fit (SM)& \multirow{2}{*}{$(0.01589,0.4352)$} & \multirow{2}{*}{1.023} & \multirow{2}{*}{-} & \multirow{2}{*}{3.16} & \multirow{2}{*}{1.0} & \multirow{2}{*}{0.3301} & \multirow{2}{*}{$4.2\times10^{-4}$}\\
(w/o $\tau_p$ limit) & & & & & & &\\
\hline
\end{tabular}
}
\caption{The parameter values for three sample cases corresponding to the light states $S$, $E$, and the SM. The best fit has a proton lifetime in conflict with the experimental limit. Therefore we also give the best fit which respects this constraint. 
In the light $S$ state case, since the best fit value for $\eta$ is large and close to the nonperturbative limit, a sample point with smaller $\eta$ is given to show that a perturbative $\eta$ value is also a viable solution.}
\label{points.TAB}
\end{table}

\begin{figure}[ht]
\centering
\includegraphics[scale=.8]{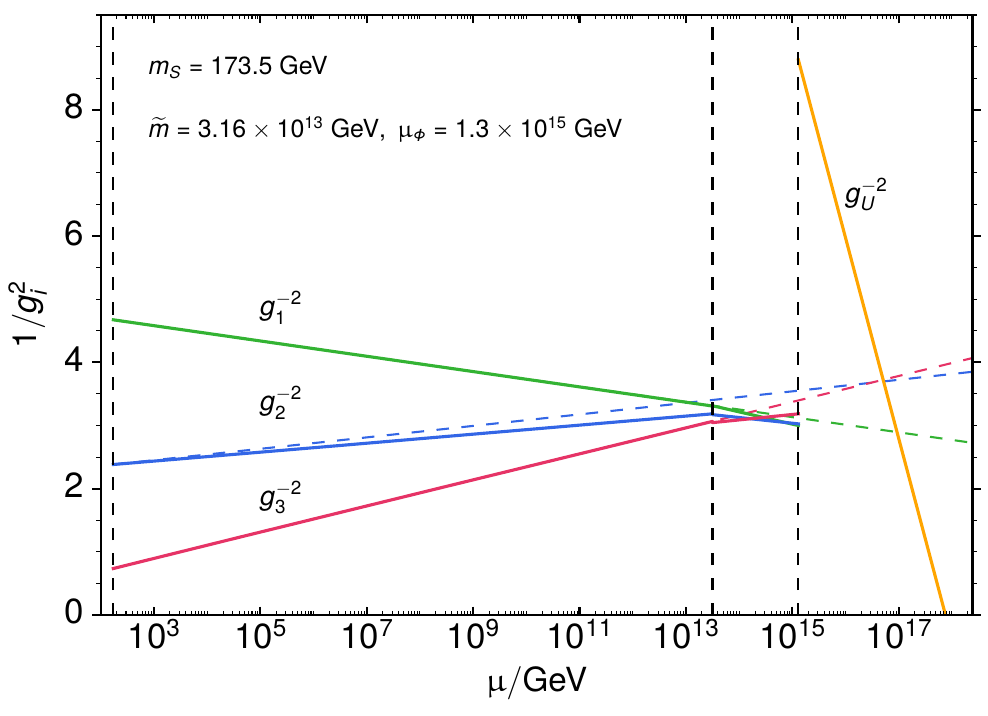}
\caption{The running of the gauge coupling are illustrated for the given parameter values. The dashed lines are the pure SM case, while the solid lines account for the inclusion of the light GUT state $S$. The parameters, $m_S, \widetilde m, \mu_\Phi$ and $g_U(\mu_\Phi)$ are taken at the best fit point, and the values are shown in Table~\ref{points.TAB}. The disparity between $g_U(\mu_\Phi)$ and $g_i(\mu_\Phi)$, is due
to approximately universal, large threshold corrections as explained in the text.}
\label{couplings.FIG}
\end{figure}

Figure~\ref{couplings.FIG} illustrates how the gauge couplings evolve and are unified into a single coupling $g_U$ when accounting for the threshold corrections for the best fit point using parameters given in Table~\ref{points.TAB}. The dashed lines in the figure show the running gauge couplings in the SM. When the renormalization scale $\mu >\mu_\Phi$, all chiral and vector multiplets participate in the running, and the RGE for $g_U$ at the 1-loop level is given by
\begin{eqnarray}
\frac{d\alpha_U^{-1}}{d\log\mu}=-\frac{b_U}{2\pi},
 \label{rge_gu.EQ}
\end{eqnarray}
where $\alpha_U\equiv g_U^2/4\pi$ and the $\beta$-function coefficient, $b_U=109$. It is particularly interesting to note that although gauge coupling unification occurs for this point,
the unified coupling $g_U$ does not match the value of the three gauge couplings at $\mu_\Phi$ due to the large threshold corrections at the GUT scale.
Though a large number of states participates in the running of $g_U$, it remains perturbative up to the scale of the heaviest state, $M_{\rm heaviest}$, due to the relatively large threshold corrections. There are regions in the $\lambda, \eta$ parameter space 
 where $g_U(M_{\rm heaviest})$ becomes much larger than unity and therefore nonperturbative. A nonperturbative bound is obtained from the condition $1/g_U^2(\mu_\Phi) < b_U/(8\pi^2)\log(M_{\rm heaviest}/\mu_\Phi)$, which corresponds to the light gray regions in Figure~\ref{chi2.FIG}. We also 
 note that with a light $S$ state, the gauge couplings, $g_i(\mu_\Phi)$ appear to focus much more 
 compared to the SM-only  case. This focusing is actually preserved by the large threshold corrections 
 $\lambda_i$, since the  corrections are very similar in magnitude. Thus the gauge couplings remain 
 unified, but at a value $g_U$  that differs from the focused value at $\mu_\Phi$, as shown in the figure.
 It should also be noted that the unified coupling becomes nonperturbative below the reduced Planck scale,
 $M_{\rm Pl}\simeq 2.4\times10^{18}$ GeV.

In the relevant figures, we have set the couplings $\alpha = \bar{\alpha} =0.58$, and let $\eta$ and $\lambda$ vary between $5\times 10^{-3}$ and $4\pi$.
This value of $\alpha = \bar{\alpha}$ is chosen to obtain the correct Higgs quartic coupling
as will be explained in section 6.  For all other states considered, we use $\alpha = \bar{\alpha}=1$.
Smaller values of $\eta$ or $\lambda$ result in one of several possible problems as seen in the 
appropriate figures.
Note that in all of our solutions, we have taken ${\widetilde m}^2 < 0$.  However,
despite the choice of a negative soft supersymmetry breaking mass-squared, there are no tachyonic states in the spectrum as the $\mu$-terms which are of order
$\mu_\Phi > |\widetilde m|$, ensure positive mass terms for all of the scalars.

For each choice of $(\lambda, \eta)$, we take $g_U, \mu_\Phi, \widetilde m$ and $m_\chi$ as free parameters, and find their values such that the function $\chi^2$ is minimized.
Figure~\ref{chi2.FIG} shows the result when the $S$ state is kept light and the white regions are free from theoretical and experimental constraints.
Note that in Figure~\ref{chi2.FIG}, we restrict the parameters to be $\widetilde m > 10^{13.5}~{\rm GeV},~\mu_\Phi > 10^{1.5} \widetilde m$ and $m_\chi > m_t$. In this figure the two thin, solid grey lines show contours of $\chi^2 = 6.18$ (lower line) and 2.3 (upper line) demarking regions corresponding to 
$> 2 \sigma$ (below the lower line), between 1 and 2$\sigma$, and $< 1 \sigma$ above the upper line. There is also a dashed line with $\chi^2 = 0.1$
and our best fit point (shown as a white circle) at large $\eta$ has $\chi^2$ close to zero (as does the solid red circle very near at slightly lower $\lambda$). 
Thus most of the viable white region has $\chi^2 < 2.3$ and is within 1$\sigma$ of  perfect unification.  
The values for the four solved parameters are given in Table \ref{points.TAB} for both the best fit (ignoring the proton lifetime constraint),
and the nearby point with sufficiently long proton lifetime.
Notice that the values of $m_{\chi}, {\widetilde m}$, and $\mu_\Phi$ are all at the edge of our prior selection.
However, relaxing these priors to $\widetilde m > 10^{13}~{\rm GeV},~\mu_\Phi > 10^{0.5} \widetilde m$ at the viable best fit point leads to an undesirable tachyonic state, which in this case, is one of the five $T$ states listed in Table \ref{MixedScalars.TAB}, since $\mu_\Phi$ is too close to $\widetilde m$. One eventually finds that $\mu_\Phi\gtrsim 10^{1.6}\widetilde m$ is needed, and our best fit point is sitting in close proximity to this boundary.

The dark gray regions hatched by diagonal lines in Figure~\ref{chi2.FIG}  are excluded by the appearance of a state with mass greater than the reduced Planck scale. In this case, there is no reason to believe our spectrum is reliable and we discard such solutions.
The light gray regions indicate that $g_U$ becomes nonperturbative below $\mu=M_{\rm heaviest}$, where $M_{\rm heaviest}$ is the mass of the heaviest state in
the spectrum. 
The dark gray regions hatched by crossed lines show the limit imposed by the proton lifetime, where the main decay channel is $p\to\pi^0e^+$ through the $(X,~Y)$ and $(X',~Y')$ gauge bosons\footnote{See, for example, Ref. \cite{Bertolini:2013vta} for more detail.}, and the current limit given by\cite{Miura:2016krn}:
\beq
\tau(p \to \pi^0 e^+) > 1.6 \times 10^{34} \text{ years} ,
\eeq
is applied. 
The proton lifetime $\tau_p \equiv \tau(p \to \pi^0 e^+)$ is proportional to $M_{X,Y}^4$ and $M_{X',Y'}^4$ with $M_{X,Y}$ and $M_{X',Y'}$ being the mass of the $(X,~Y)$ and $(X',~Y')$ gauge bosons, respectively.
Since those masses are proportional to $\lambda^{-1}$ as shown in Table~\ref{GaugeBosons.TAB}, smaller values of $\lambda$ leads to a longer $\tau_p$.
Note that around the red circle in the figure, although a larger $\eta$ may lead to a smaller $M_{X',Y'}$, the proton lifetime becomes longer in this parameter region, since the scale of $\mu_\Phi$ becomes larger when minimizing $\chi^2$, and thus the proton decay limit is relaxed.
It should also be noted that for $|\widetilde m| \gtrsim 10^{10}$ GeV the proton decay induced by dimension five operators is sufficiently suppressed.
However, the decay channel $p\to\pi^0e^+$ is also induced by a color triplet Higgs boson, and the lifetime is given by
\beq
\tau(p\to\pi^0 e^+) \simeq2\times10^{34}~{\rm years}\times\left(\frac{M_T}{1.2\times10^{11}~{\rm GeV}}\right)^4,
\eeq
where $M_T$ is the mass of the lightest color triplet Higgs boson in $H, \Sigma, \overline{\Sigma}$ and $\Phi$, and we have assumed that the Yukawa couplings are 
the same as in the SM.
In the parameter space presented in Figure~\ref{chi2.FIG}, we find that the limit on $M_T$ from proton decay, overlaps with the other constraints considered, and thus it is not shown explicitly in the figure.

\subsection{Best fit point}
\label{Best fit point}
\begin{figure}[ht!]
\centering
\includegraphics[scale=.6]{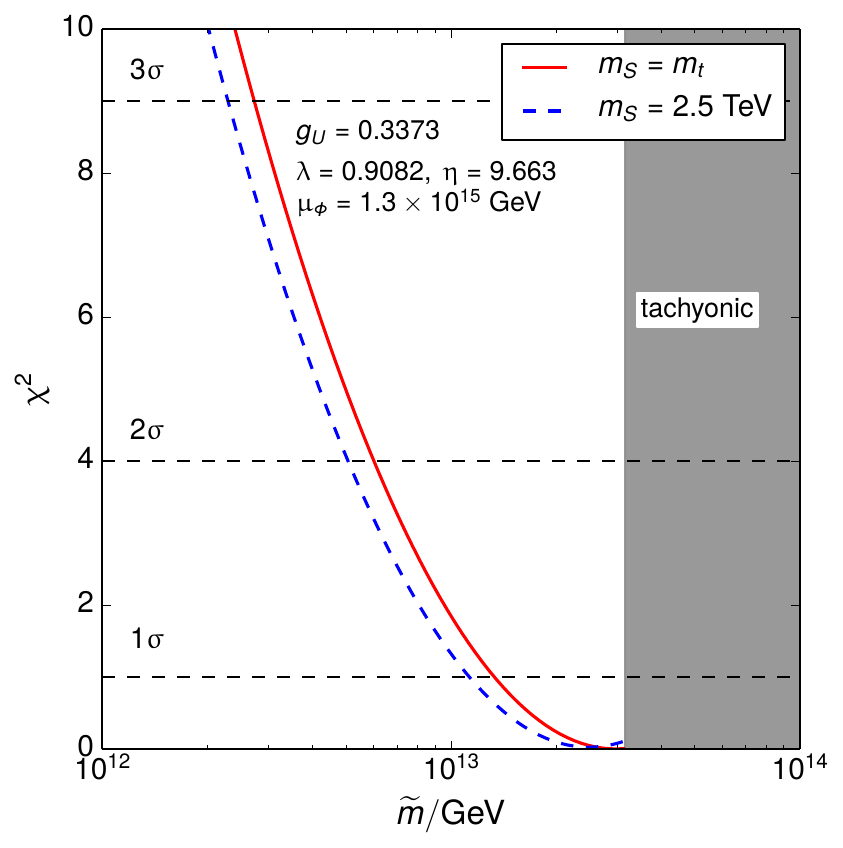}
\includegraphics[scale=.6]{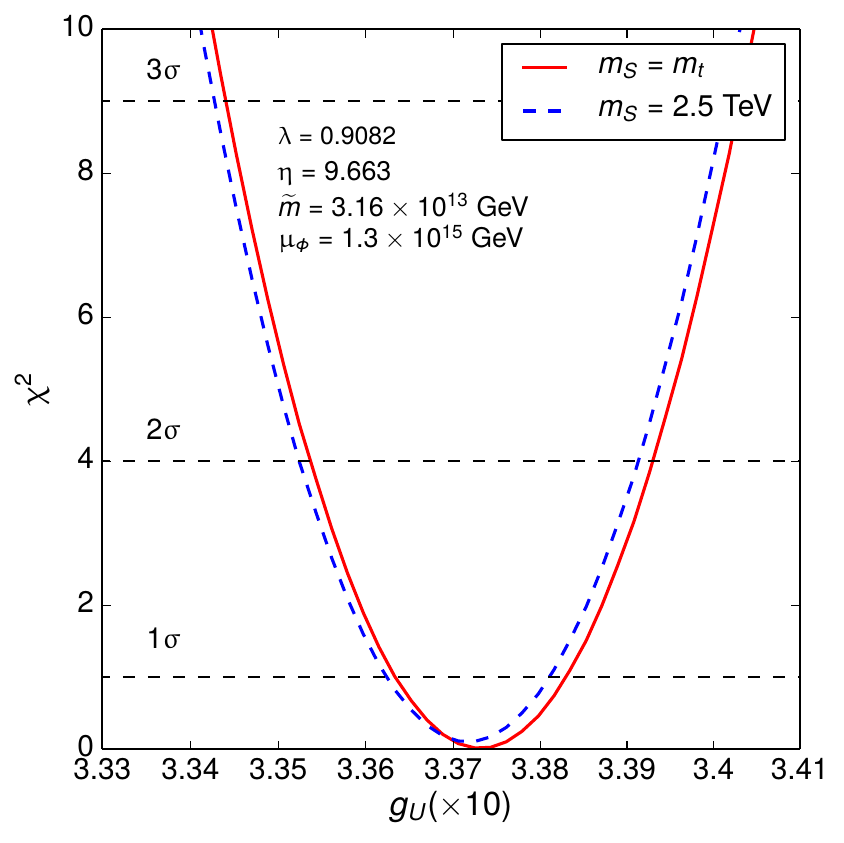}
\caption{The left (right) panel shows the value of $\chi^2$ as a function of $\widetilde m$ ($g_U$) around the best fit point. The red solid and blue dashed lines depict the case of $m_S=m_t$ and $2.5~{\rm TeV}$, respectively. The gray region in the left panel is excluded due to the presence of an undesirable tachyonic state.}
\label{chi2_msusy_gu.FIG}
\end{figure}

We now take a closer look at our viable best fit point in Figure~\ref{chi2.FIG} by varying the relevant parameters.
The left (right) panel of Figure~\ref{chi2_msusy_gu.FIG} shows $\chi^2$ as a function of $\widetilde m$ ($g_U$) with fixed $\lambda,~\eta,~g_U$ and $\mu_\Phi$ ($\lambda,~\eta,~\widetilde m$ and $\mu_\Phi$).
In both panels, $m_S$ is taken to be $m_S=173.5~{\rm GeV}$ and $2.5~{\rm TeV}$, and as one sees $\chi^2$ is relatively flat along the variation of $m_S$, indicating that larger $m_S$ will give a similar $\chi^2$ with the rest of the parameters being the same.
In the left panel of Figure~\ref{chi2_msusy_gu.FIG}, the gray shaded region is excluded due to the presence of an undesirable tachyonic state which breaks the SM gauge symmetry\footnote{One of the $T$ states becomes tachyonic in this case.}.
We note that one should not interpret these figures as providing the true uncertainty in either ${\widetilde m}$ or $g_U$.  We have held fixed the remaining parameters rather than having
marginalized over them, and we expect that as a function of either ${\widetilde m}$ or $g_U$, allowing the remaining parameters to vary freely, $\chi^2$ would become
quite a bit flatter allowing a broader range in ${\widetilde m}$ or $g_U$. Nevertheless, these curves give us an idea of the shape of the $\chi^2$ function in certain directions
of parameter space.

\begin{figure}[ht!]
\centering
\includegraphics[scale=.6]{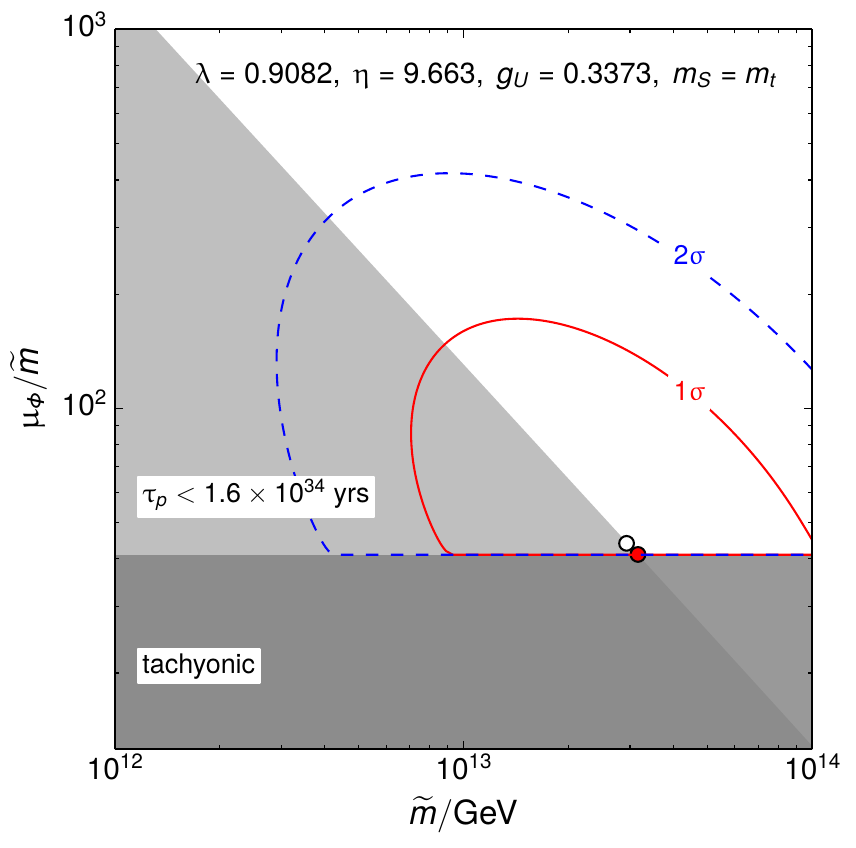}
\includegraphics[scale=.6]{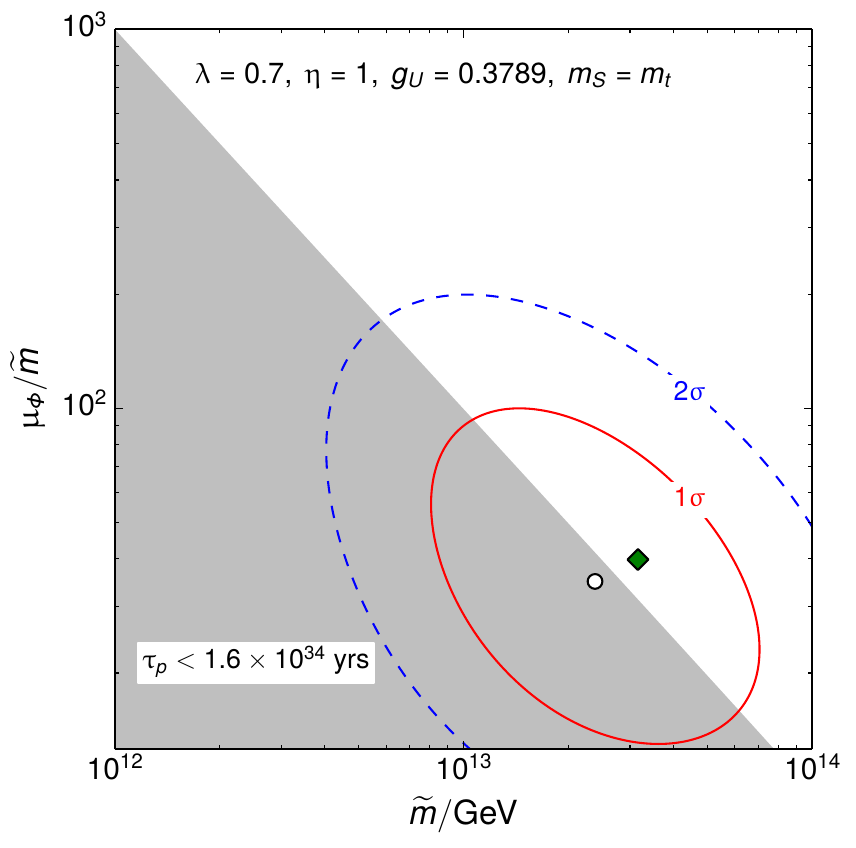}
\caption{The $1\sigma$ and $2\sigma$ regions are shown by taking $\widetilde m$ and $\mu_\Phi$ as free parameters. The light and dark gray regions are excluded by the proton lifetime limit and the presence of an undesirable tachyonic state (only in the left panel), respectively. In the left panel, the red circle is our best fit point, while the white circle indicates the best fit point without the proton decay constraint. In the right panel, the green diamond represents the sample point listed in Table~\ref{points.TAB}, while the white circle shows the best fit point without the proton decay constraint.}
\label{chi2_msusy_muPhi.FIG}
\end{figure}

\begin{figure}[ht]
\centering
\includegraphics[scale=.6]{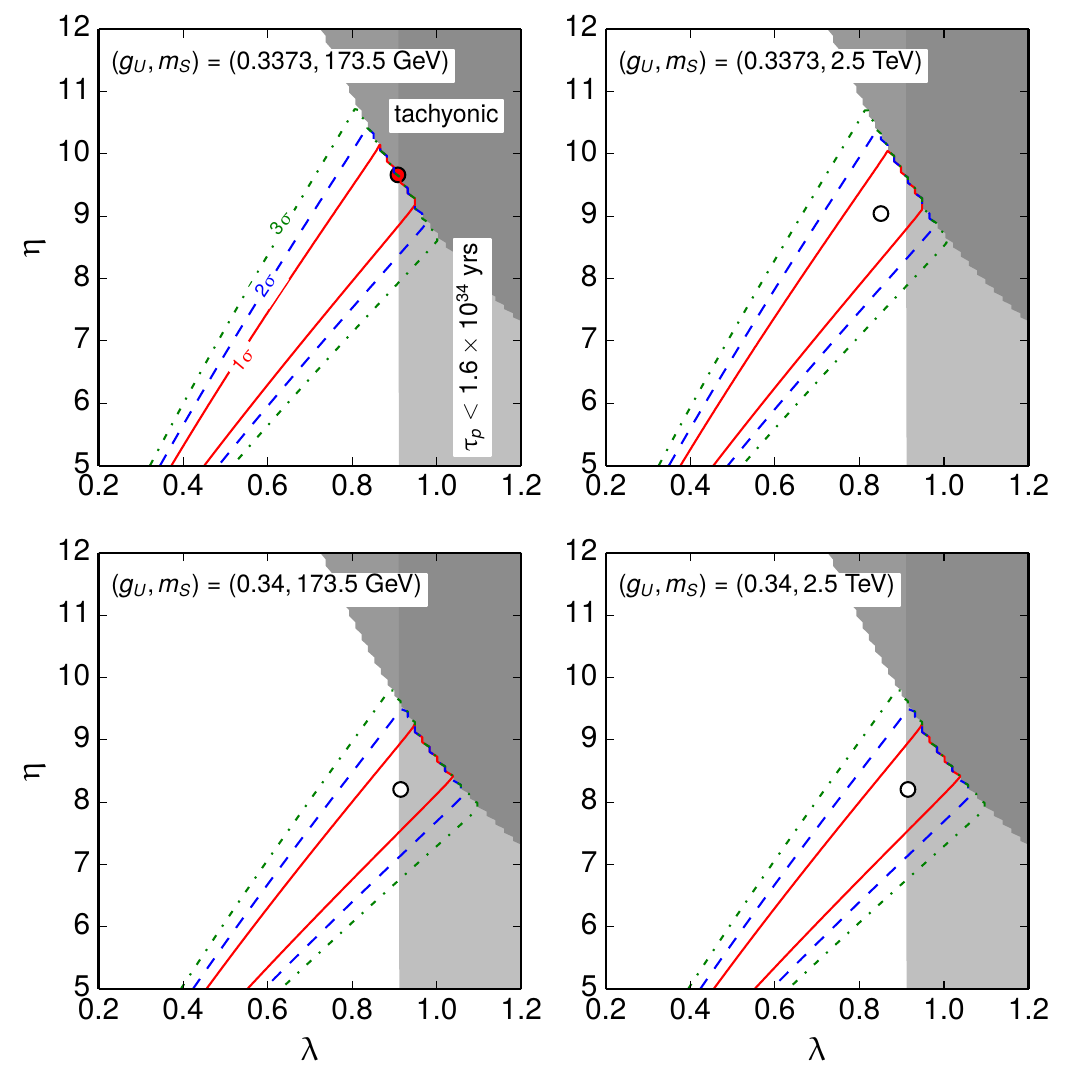}
\caption{The best fit points for $(g_U,~m_S)=(0.3373,~173.5~{\rm GeV}),~(0.3373,~2.5~{\rm TeV}),(0.34,~173.5~{\rm GeV}),~(0.34,~2.5~{\rm TeV})$ are shown. The red circle in the upper-left panel is our viable best fit point, while the white circles in the figure correspond to the best fit points without the proton decay limit. The light and dark gray regions are excluded by the proton lifetime limit and the appearance of a tachyonic state, respectively. The red solid, blue dashed and green dot-dashed lines represent the $1\sigma$, $2\sigma$ and $3\sigma$ regions, respectively. The rest of the parameters are taken to be the same as for our best fit point, i.e., $\widetilde m=3.16\times10^{13}~{\rm GeV},~\mu_\Phi=1.3\times10^{15}~{\rm GeV}$.}
\label{gu_ms.FIG}
\end{figure}

Figure~\ref{chi2_msusy_muPhi.FIG} shows the $1\sigma$ and $2\sigma$ regions in the $(\widetilde m,~\mu_\Phi/\widetilde m)$ plane, where the light and dark gray regions are excluded by the proton lifetime limit and the presence of a tachyonic state (only in the left panel), respectively. 
The left panel of the figure shows the best fit point, while for comparison, the right panel shows the sample point listed in Table~\ref{points.TAB}.
Once again, the remaining parameters are held fixed.
Concerning the presence of a tachyonic state, although there is no such parameter region in Figure~\ref{chi2.FIG}, it appears in the left panel of Figure~\ref{chi2_msusy_muPhi.FIG} as we have fixed $\mu_\Phi$.
When we take larger $\mu_\Phi$, none of the GUT Higgs fields become tachyonic.
Our viable best fit point is indicated by the red circle, while the best fit point without the proton decay constraint is depicted by the white circle. 
In the right panel of Figure~\ref{chi2_msusy_muPhi.FIG}, the sample point is indicated by the green diamond, while the true best fit point for this choice of $\lambda$ and $\eta$ appears as the white circle which is inside the proton decay constraint.
Note that a tachyonic state does not show up for the plotted range of $\mu_\Phi$ due to the smaller value of $\eta$.
We also exhibit how the best fit point moves by varying $g_U$ and $m_S$ in Figure~\ref{gu_ms.FIG}, where the upper-left panel includes our viable best fit point.
The meaning of the white circles are the same as in the previous figures.
The dark gray region is excluded by the appearance of a tachyonic state.
Note that in the upper-left panel, there is no white circle, and the red circle corresponds to the true best fit point since we have fixed $\widetilde m=3.16\times10^{13}~{\rm GeV}$ and $\mu_\Phi=1.3\times10^{15}~{\rm GeV}$, and the dark gray region eliminates the smaller $\chi^2$ region. 
While the best fit point is not sensitive to $m_S$ as expected, larger $g_U$ moves the point toward the smaller $\eta$, which indicates that the threshold corrections also become smaller as $g_U^{-2}$ and $\lambda_i/(48\pi^2)$ should be balanced to get $\chi^2$ smaller.
In the smaller $\eta$ region, on the other hand, the larger $g_U$ easily reaches the nonperturbative region as indicated in Figure~\ref{chi2.FIG}.
With the exception of the proton lifetime constraint and the appearance of a tachyonic state, we have suppressed the other constraints seen in Figure~\ref{chi2.FIG} for clarity.

Finally, we note that if $S$ is the lightest state, it is very long-lived and hence becomes another (in addition to the gravitino) viable dark matter candidate.
The relative stability of $S$ is ensured by the absence of any linear and cubic terms for $S$ due to SU(2)$_L$.\footnote{For instance, the cubic terms vanish because ${\rm tr}[\sigma^a\sigma^b\sigma^c]=2i\epsilon^{abc}$, where $\sigma^a$ are the Pauli matrices.} Instead, the coupling $S\widetilde h\widetilde h$ arising from $\Phi\Sigma\overline\Sigma\supset (3,1,15)\times(2,2,15)\times(2,2,15)$ may induce the decay of $S \to \gamma \gamma$ at one loop since the Higgsinos $\widetilde h$ are much heavier than $S$. Its decay however, is greatly suppressed due to the large mass difference 
between $m_S \sim \mathcal{O}(1)$ TeV and $m_{\widetilde h} \simeq {\widetilde m}$. 
The phenomenology of hypercharge-zero, scalar triplets has been considered widely in the literature (see e.g. \cite{Ross:1975fq, Cirelli:2005uq, Cirelli:2007xd, Cirelli:2008id, FileviezPerez:2008bj}). 
In fact, such a scalar is known as one of the minimal dark matter candidates \cite{Cirelli:2005uq,Cirelli:2007xd}, and the upper bound on $m_S$ is given by $m_S\lesssim 2.5~{\rm TeV}$ by demanding its present relic density should be smaller than the observed value~\cite{Cirelli:2007xd}.
However, in the majority of previous studies, the triplet is taken to be real, which is not the case in our construction. 
For a complex scalar triplet with zero hypercharge, we expect that the observed relic density would be obtained for a scalar mass of a similar order. 
In what follows, we also keep this upper bound in mind. A more detailed phenomenological study of $S$ and its potential for observation of 
TeV $\gamma$-rays will be treated elsewhere.

\begin{figure}[ht!]
\centering
\includegraphics[scale=0.45]{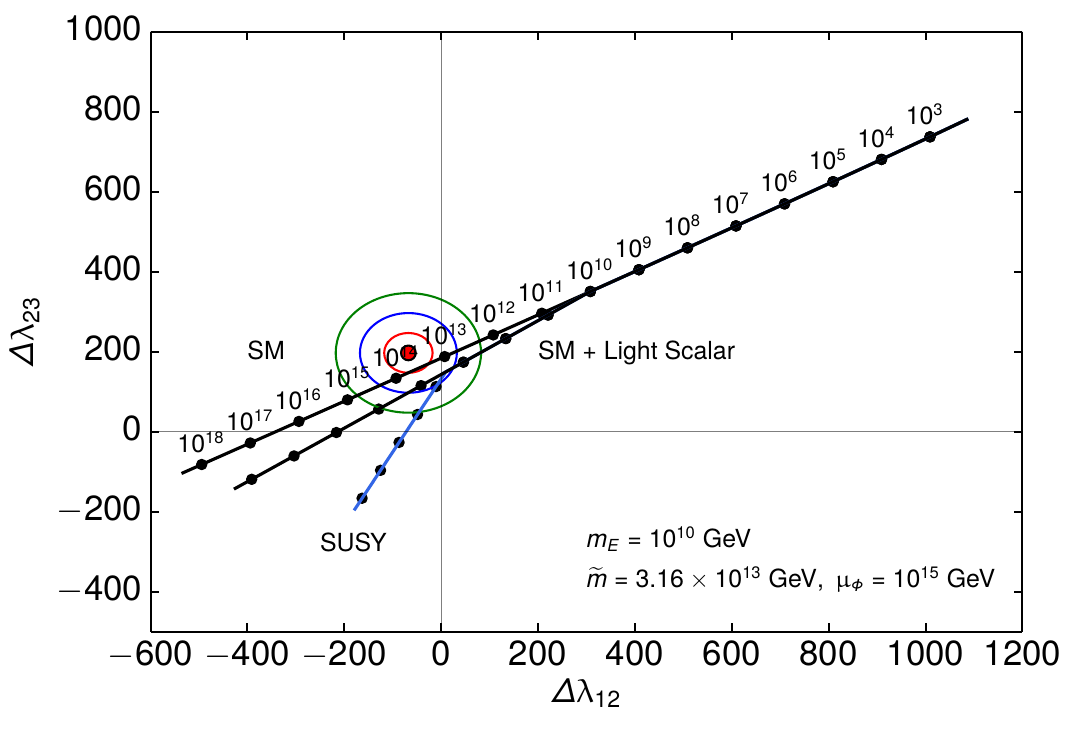}
\includegraphics[scale=0.45]{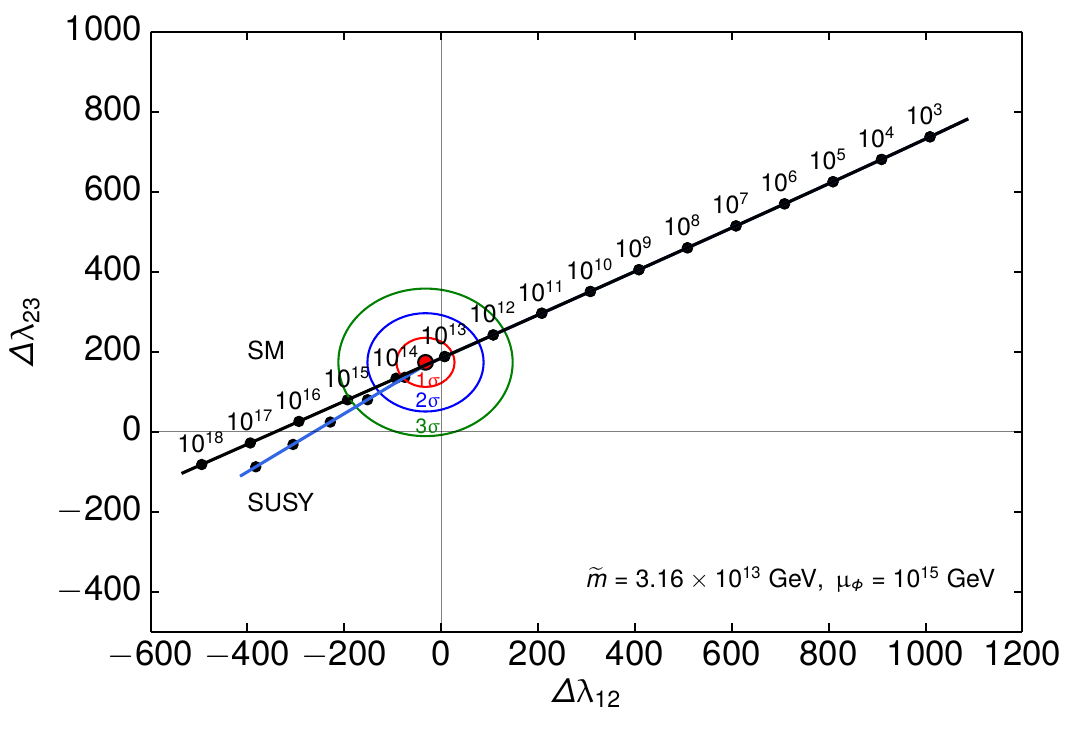}
\caption{$\Delta\lambda$ plot for $E$ ($\Sigma(\bar{3},2,-1/6)+$ h.c.) as the only light state (left), and only the SM below $\widetilde{m}$ (right).}
\label{130_Delta_Lambda_10to4e.FIG}
\end{figure}

\begin{figure}[ht]
\centering
\includegraphics[scale=0.6]{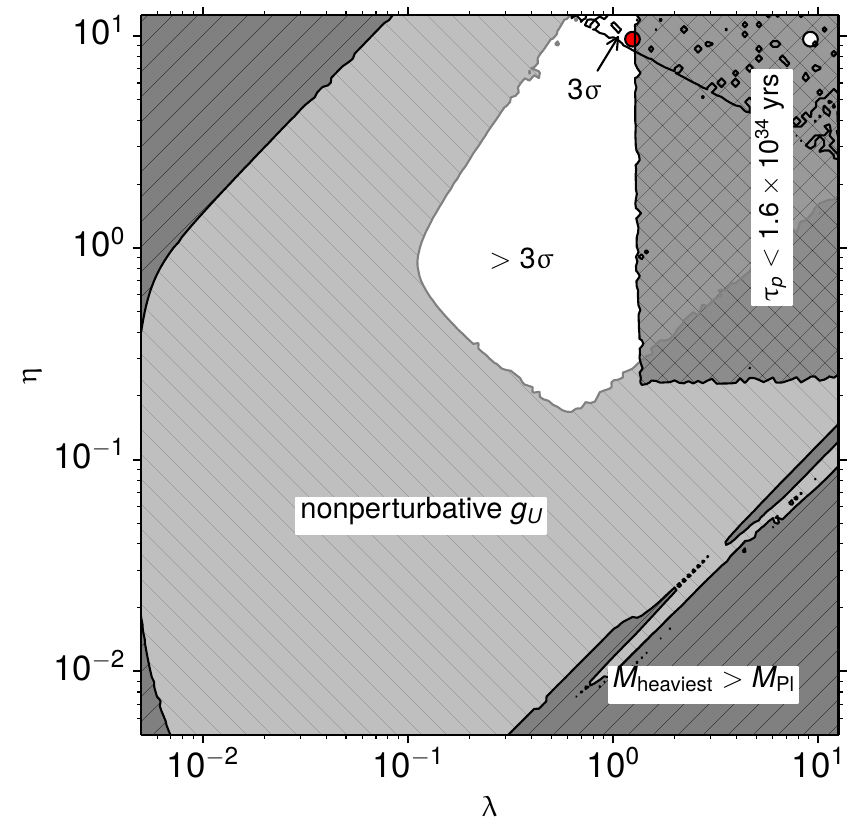}
\includegraphics[scale=0.6]{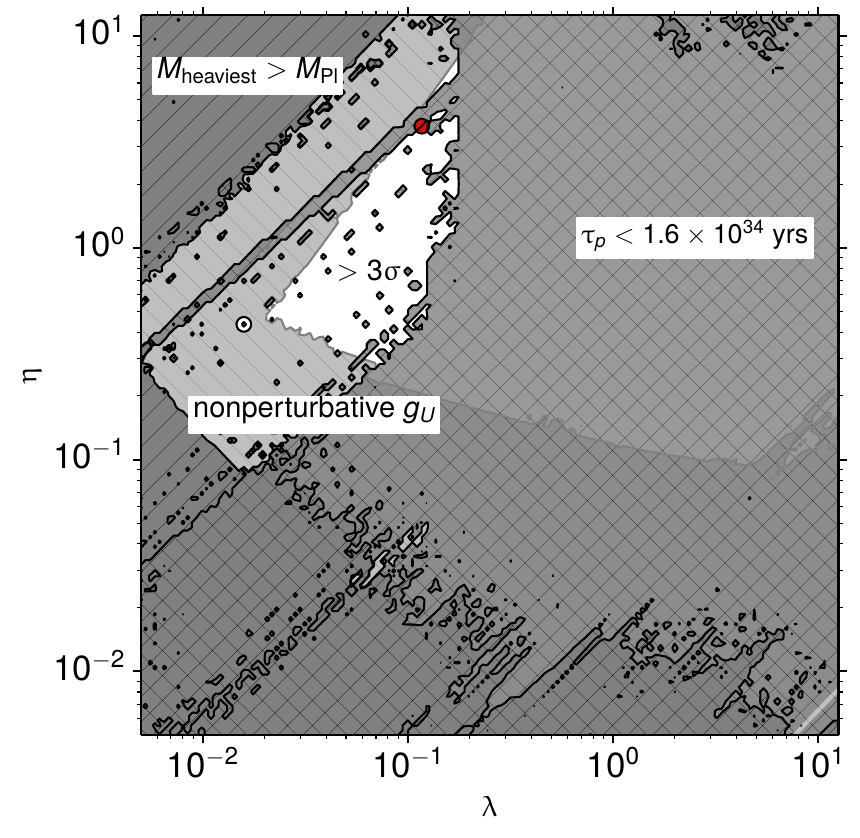}
\caption{As in Figure~\ref{chi2.FIG}, viable regions for the coupling unification are indicated by the white region. The left (right) panel of the figure shows the case of the light state $E$ (SM). 
The white (red) circles indicate the (viable) best fit points whose detailed parameters are listed in Table~\ref{points.TAB}.}
\label{chi2e.FIG}
\end{figure}

We have also tested for unification solutions for the other states listed in Table \ref{UnmixedScalars.TAB} as well as the $G$ state listed in Table \ref{MixedScalars.TAB} by taking $\alpha=\bar \alpha=1$.
Surprisingly, despite the large number of potential candidates, no reasonable solutions were found. The $E$ state ($\Sigma$($\bar{3},2, -1/6)+$ h.c.) was the second best candidate, with all others yielding higher values of $\chi^2$.
For comparison, we show in Figures~\ref{130_Delta_Lambda_10to4e.FIG} and \ref{chi2e.FIG} results for the $E$ state and the SM.
The $\Delta \lambda$ planes for $E$ and the SM model are shown in Figure~\ref{130_Delta_Lambda_10to4e.FIG}.
As one can see from the left panel, the best viable point for the $E$ state lies well off the SUSY line with $\mu = 10^{15}$ GeV. 
In the right panel for the SM, although the point lies on the line, unification would require it to sit on the line at $\mu = 10^{15} $ GeV.
Both solutions are acceptable at the 3$\sigma$ level.  Figure \ref{chi2e.FIG} shows the corresponding $(\lambda, \eta)$ planes.
In both cases, there are regions where all constraints considered are viable, however unification at the best fit point occurs at no better
than the 3$\sigma$ level\footnote{We remind the reader that the meaning of $\sigma$ in Figures \ref{130_Delta_Lambda_10to4e.FIG} and \ref{chi2e.FIG}
is not exactly the same. In Figure~\ref{130_Delta_Lambda_10to4e.FIG}, $\sigma_{\Delta \lambda_{ij}} = 0.01 (48 \pi^2) (\hat{\lambda}_i^2 
+ \hat{\lambda}_j^2)^{1/2}$ as discussed above and determines the size of the ellipses, whereas in Figure~\ref{chi2e.FIG},
$\sigma$ is determined from the value of $\chi^2$.}. In order for $E$ to remain light, the value of $x$ in the SUSY limit is $x = 0.6133 + 0.7339 i$.
As one can see from the figure and Table \ref{points.TAB}, the true best fit point (white circle) lies at higher $\lambda$ than the best viable point
(red circle) with an acceptable proton lifetime. In the right panel of Figure~\ref{chi2e.FIG} for the SM, there is no unique value for $x$
as no state is tuned to be light. Instead, we have scanned over $x$ and in fact at each point shown there is a different value of $x$ (and 
hence the heavy particle spectrum) which minimizes $\chi^2$. As a result a best fit point (white circle) is found with $x = -0.9955$.
In fact as one can see from Table \ref{points.TAB}, the value of $\chi^2$ at this point is of order 1\footnote{Because the value of $x$ differs at each point,
the $\chi^2$ boundary is complicated and contains islands as seen in the figure.}. However at this particular value of $x$,
the proton lifetime is far too small. The best viable point (red circle) has $x = -0.6828$, but has a significantly larger value of $\chi^2$. 
Of course the SM on its own can not resolve the Higgs stability question discussed in the next section.

Finally, we mention one additional case of potential interest, that of a SM singlet, $G$ which remains light if $x = -0.9796$ (corresponding to $x=-1$ in the SUSY-preserving limit) at the best fit point.
In principle one might expect this case to be as good as the SM, while still providing a possible solution to the Higgs vacuum stability by virtue of $G$ coupling to the SM Higgs. However there are no solutions
in the $(\lambda, \eta)$ plane with a small $\chi^2$ value equivalent to a significance less than 3$\sigma$, and since $x$ is fixed, the spectrum cannot be adjusted to obtain a better fit. Furthermore the singlet $G$ case is plagued by proton decay constraints since the GUT gauge bosons are light.
Thus in the following discussion on vacuum stability, we will restrict ourselves to the case of the light $S$ state, and denote $m_\chi = m_S$.

\section{Electroweak vacuum stability}
\label{Electroweak vacuum stability}

As we have seen, gauge coupling unification is in general possible even when all the Higgs multiplets are heavy, and 
the supersymmetry breaking scale is large.  However, in order to correct the running of the low energy Higgs quartic coupling,
we must require some deviation from the SM at energies below roughly $10^{10}$ GeV. In this section,
we concentrate on the unification model with a light state $S$.

The light SU(2)$_L$ triplet scalar couples to the SM Higgs field, and its effect on RGE evolution may keep the Higgs quartic coupling from running negative.
The relevant part of the scalar potential is given by
\begin{eqnarray}
V(H,S) &=& V_F + V_D + V_{\rm MSSM}(H_u,H_d),\\
V_F &=& \frac{1}{2}(|\alpha|^2+|\bar\alpha|^2)(S^aS^{a*})(|H_u|^2+|H_d|^2)+\frac{17}{3}|\lambda|^2(S^aS^a)(S^bS^b)^*,\\
V_D&=& -\frac{1}{2}g_2^2[(S^aS^a)(S^bS^b)^*-(S^aS^{a*})^2] \, .
\end{eqnarray}
Note that because the SM Higgs doublets are actually linear combination of doublets in the {\bf 10}, {\bf 126}, ${\bf \overline{126}}$, and {\bf 210},
the full $F$-term coupling of $S$ to $H_{u,d}$ is significantly more complicated. However, we have checked explicitly for the best fit point, that the components
of the {\bf 126}, ${\bf \overline{126}}$, and {\bf 210} in $H_{u,d}$ are extremely small and we can approximate $H_{u,d}$ as being derived solely from the {\bf 10}.\footnote{Note that this remains a good approximation as long as $\alpha/\lambda$, $\overline\alpha/\lambda$, $\alpha/\eta$, and $\overline\alpha/\eta$ are smaller than ${\cal O}(1-10)$, and for larger values for those ratios, the fraction of the components of the {\bf 126}, ${\bf \overline{126}}$, and {\bf 210} in $H_{u,d}$ increases as the off-diagonal entries of ${\cal B}^h$ given by Eq. (\ref{eq:Bh}) turn out to be proportional to those ratios.}
The heavy components of the {\bf 10} should be integrated out below ${\widetilde m}$. 
In order to work in a basis which will simplify the connection to the phenomenology in the IR, we perform a rotation on the Higgs doublets to the so-called ``Higgs basis"
\cite{hssusy2}, 
\beq
\bmat H \\ A \emat = \bmat \cos\beta & \sin\beta \\ -\sin\beta & \cos\beta \emat \bmat -\epsilon H_d^* \\ H_u\emat \ ,
\eeq
where $H$ is identified as the light Higgs doublet, while $A$ remains at the SUSY scale. The rotation angle $\beta$ 
differs from the usual $\tan\beta = v_u / v_d$ by  $\mathcal{O}(m_Z^2/\widetilde{m}^2)$ in an appropriately chosen scheme. 
This will simplify the matching between the broken and unbroken SUSY phases. 

The scalar potential of the light Higgs and scalar triplet below the SUSY scale $\widetilde{m}$ can be written as
\begin{align}
 V(H,S) &\supset m_H^2 |H|^2  + m_S^2\big(S^a {S^a}^*\big)   + \lambda_H |H|^4 + \lambda_{HS} \big(S^a {S^a}^*\big) |H|^2 + \lambda_S \big(S^a {S^a}^*\big)^2 + \lambda_{SS^*} \big( S^a S^a\big)\big(S^b S^b\big)^*  \ ,
\end{align}
where $m_H$ should not be interpreted as the physical Higgs boson mass, $m_h$, but rather as the potential mass parameter, which after electroweak symmetry breaking (EWSB) is related to the Higgs mass by $m_h^2 = 2 \lambda_H v^2 = - m_H^2$. 
The matching conditions between the broken and unbroken SUSY phases are
\begin{align}
 \nonumber
 \lambda_H (\widetilde{m}) &= \frac{1}{8}\left(\frac{3}{5}g_1^2(\widetilde{m}) + g_2^2(\widetilde{m}) \right) \cos^2 2\beta \ ,~~~
 \lambda_{HS} (\widetilde{m}) = \frac{1}{2}\left( |\alpha|^2+| \bar{\alpha}|^2\right) \ , \\
 \lambda_S (\widetilde{m}) &= \frac{1}{2}g_2^2(\widetilde{m}) \ , ~~~~~~~~~~~~~~~~~~~~~~~~~~~~~~~
 \lambda_{SS^*} (\widetilde{m}) = \frac{17}{3}|\lambda|^2 - \frac{1}{2}g_2^2(\widetilde{m}) \ ,
\end{align}
where we have assumed an approximately degenerate SUSY spectrum so that the one-loop threshold corrections (given in \cite{hssusy2}) can be ignored. 
The matching conditions are in effect boundary conditions for the RGE running of the quartic couplings. 
Thus, we must check that there are solutions to the RGEs satisfying the boundary conditions at ${\widetilde m}$, which we take for definitiveness to be $3 \times 10^{13}$ GeV, and the weak scale (or $m_t$).
This is, in fact, non-trivial, as there are few adjustable parameters at our disposal: $\lambda, \alpha={\bar \alpha}, m_S$, and  $\tan \beta$. 

In the broken SUSY phase, the RGE for the Higgs quartic coupling and mass term can be found for example in \cite{Buttazzo:2013uya}, and at one-loop level they are given by\footnote{We define $dx/ dt = \beta_x^{(1)}$ where $t= \log \mu$ as opposed to $t = \log \mu^2$ in \cite{Buttazzo:2013uya}.}
\begin{align}
(4\pi)^2 \beta^{(1)}_{\lambda_H} =& 24\lambda_H^2-\lambda_H\left(\frac{9}{5}g_1^2+9g_2^2\right)+\frac{3}{4}g_2^4+\frac{3}{8}\left(\frac{3}{5}g_1^2+g_2^2\right)^2 -6y_t^4+12\lambda_Hy_t^2,\\
(4\pi)^2 \beta^{(1)}_{m_H^2} =& \left(12\lambda_H-\frac{9}{10}g_1^2 -\frac{9}{2}g_2^2 + 6y_t^2\right) m_H^2 \ . 
\end{align}
There is however a modification due to the inclusion of the operator coupling $H$ to $S$ with coupling $\lambda_{HS}$. At one loop, this modification is given by
\begin{align}
(4\pi)^2 \delta\beta^{(1)}_{\lambda_H} =~& 3\lambda_{HS}^2\ , \\
(4\pi)^2 \delta\beta^{(1)}_{m_H^2} =~& 6\lambda_{HS} m_S^2 \ ,
\end{align}
which should be added to the usual one-loop $\beta$-function coefficient for the Higgs quartic coupling $\beta^{(1)}_{\lambda_H}$ and the Higgs quadratic term $\beta^{(1)}_{m_H^2}$, respectively.

\begin{figure}[ht!]
\centering
\includegraphics[scale=0.7]{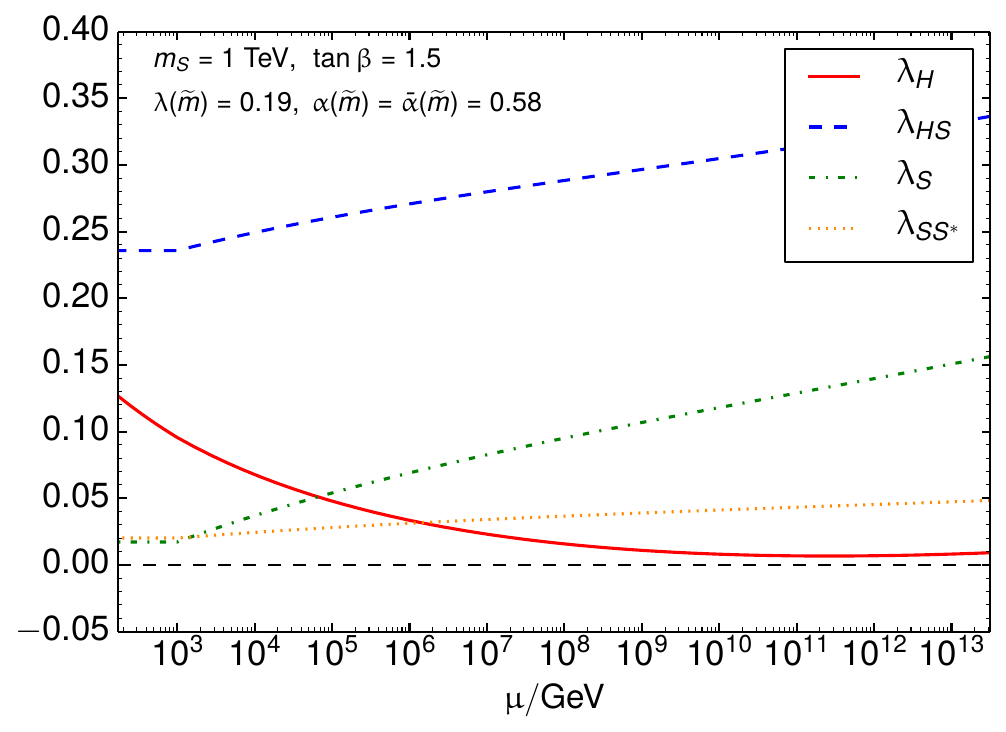}
\caption{The renormalization group running of each quartic coupling for the given fixed parameters. 
The black dashed line indicates the zero of the vertical axis. The red line, showing the running of $\lambda_H$, never goes below zero, ensuring a stable Higgs potential. The boundary value of $\lambda_H$ at $m_t$ is seen
to agree with $\lambda_H^{\rm exp}(m_t)$.}
\label{lambdaH1.FIG}
\end{figure}

The RGEs for the new scalar potential couplings at one loop are
\begin{align}
(4\pi)^2 \beta^{(1)}_{m_S^2} =~& 4\lambda_{HS} m_H^2 + (16 \lambda_S + 8 \lambda_{SS^*} - 12g_2^2) m_S^2\ , \\
 (4\pi)^2 \beta^{(1)}_{\lambda_{HS}} =~& 4 \lambda_{HS}^2 + 12 \lambda_H \lambda_{HS} + 8 \lambda_{HS} \lambda_{SS^*} +  16 \lambda_{HS} \lambda_{S}  
+ 6 \lambda_{HS} y_t^2 -\frac{9}{10} g_1^2 \lambda_{HS} - \frac{33}{2} g_2^2 \lambda_{HS} + 6 g_2^4\ , \\
(4\pi)^2 \beta^{(1)}_{\lambda_{S}} =~& 28 \lambda_{S}^2+ 2 \lambda_{HS}^2 + 16 \lambda_{SS^*}^2 +  16 \lambda_{SS^*} \lambda_{S}  -24 g_2^2 \lambda_{S} + 9 g_2^4\ , \\
(4\pi)^2 \beta^{(1)}_{\lambda_{SS^*}} =~& 12 \lambda_{SS^*}^2 +  24 \lambda_{SS^*} \lambda_{S}  -24 g_2^2 \lambda_{SS^*} + 3 g_2^4\, .
\end{align}
With the exception of the $g_1^2\lambda_{HS}$ term in $\beta^{(1)}_{\lambda_{HS}}$, these are consistent with the RGEs for type-II seesaw models that have a triplet Higgs field charged under U(1)$_Y$ \cite{TripletRGEs}.
For the boundary condition of $\lambda_H$ at low energy, we take $\lambda_H^{\rm exp}(\mu=m_t)=0.1261\pm0.0007$ \cite{Buttazzo:2013uya} where the uncertainties in $m_h$ and $m_t$ given in Table~\ref{tab:parameters} are taken into account.
Figure~\ref{lambdaH1.FIG} shows one of the viable cases for the vacuum stability, where $m_S=1~{\rm TeV},~\tan\beta=1.5,~\lambda(\widetilde m)=0.19,~\alpha(\widetilde m)=\bar\alpha(\widetilde m)=0.58$.
The red solid, blue dashed, green dot-dashed, and orange dotted lines are the running of $\lambda_H,~\lambda_{HS},~\lambda_S,$ and $\lambda_{SS^*}$, respectively.
The horizontal black dashed line indicates the zero of the vertical axis, and $\lambda_H$ never goes below this line.

\begin{figure}[ht!]
\centering
\includegraphics[scale=0.4]{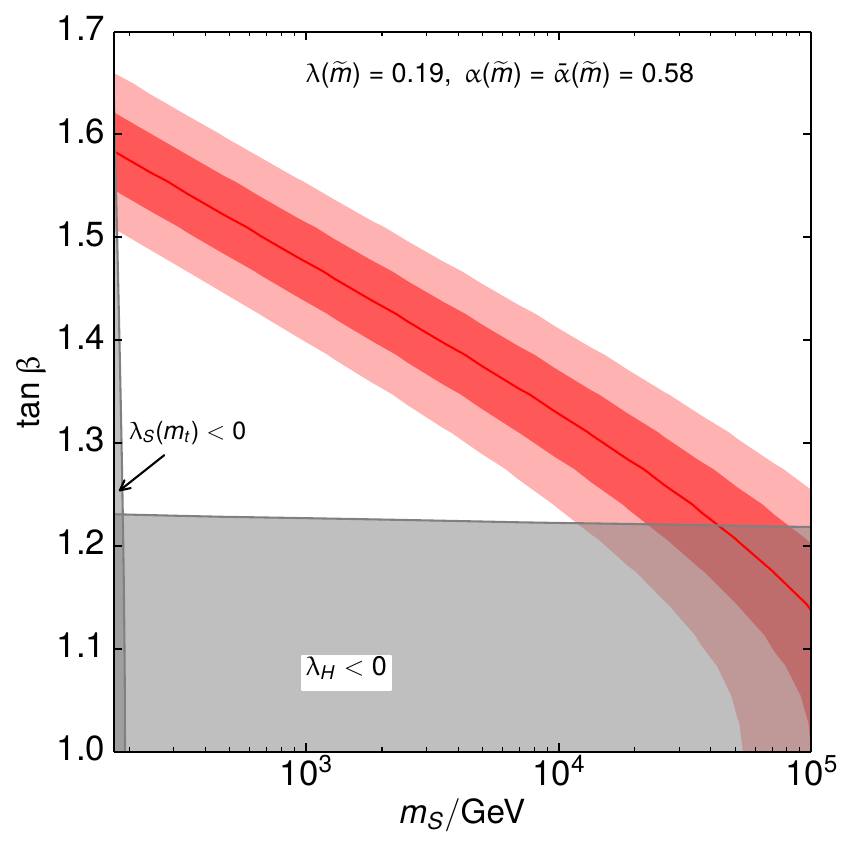}
\includegraphics[scale=0.4]{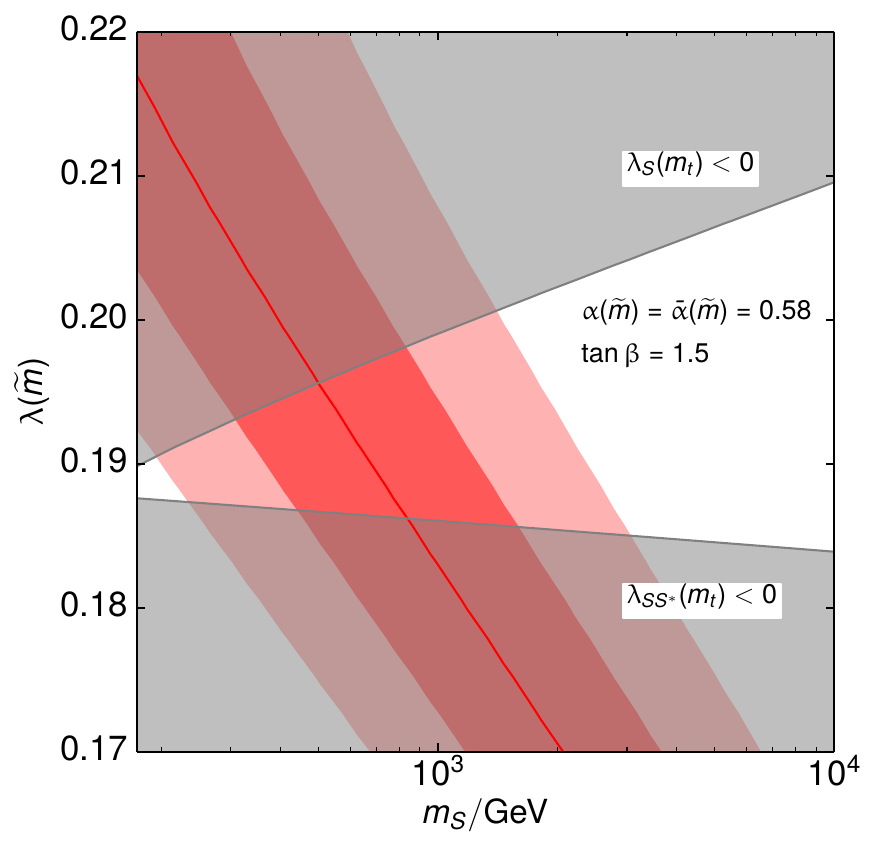}
\includegraphics[scale=0.4]{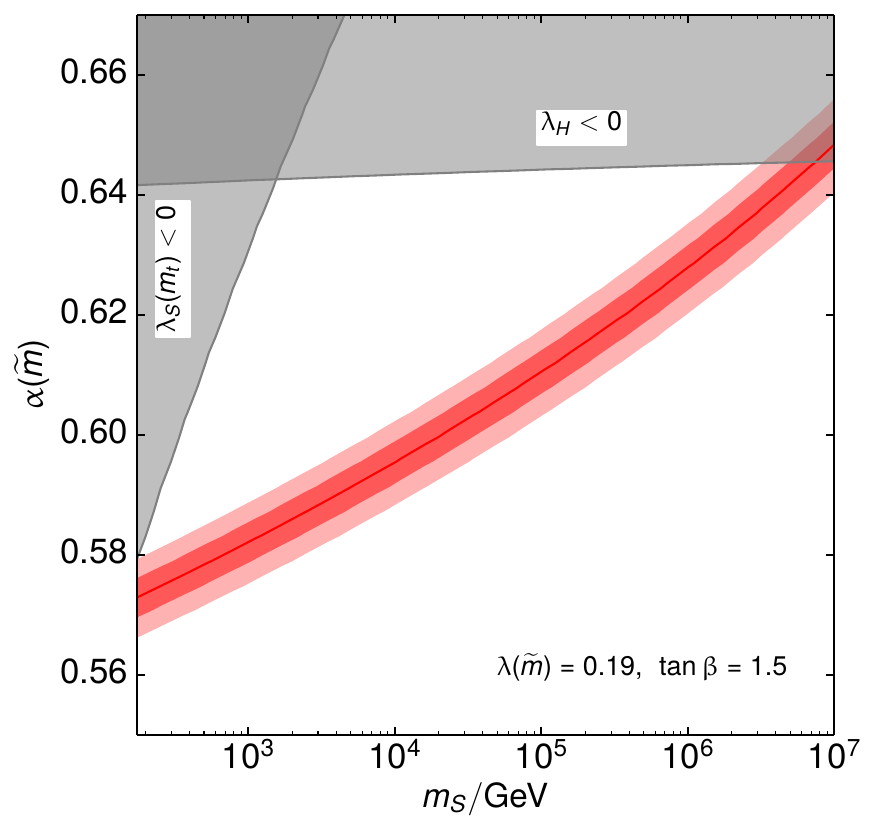}
\caption{The sensitivity of the viable region to $\tan\beta$ (left), $\lambda(\widetilde m)$ (middle), and $\alpha(\widetilde m)=\bar\alpha(\widetilde m)$ (right). The red line shows the central value of $\lambda_H^{\rm exp}(m_t)$, and the dark and light pink regions represent the $1\sigma$ and $2\sigma$ range, respectively. The gray regions are excluded due to any of $\lambda_H$, $\lambda_S$, and $\lambda_{SS^*}$ running negative.}
\label{lambdaH2.FIG}
\end{figure}

The renormalization group evolution of the quartic couplings is sensitive to the boundary values of the couplings, $\tan\beta$, and $m_S$. Of particular concern is the value of $\lambda_H(m_t)$, and the fact that $\lambda_H$ (and indeed all of the quartic couplings) remain positive and perturbative over the 
renormalization scale range of $m_t$ to ${\widetilde m}$. Generally, the values of $\alpha={\bar \alpha}$ at $\mu = {\widetilde m}$ has a strong effect on $\lambda_H(m_t)$
and so those values are adjusted to give $\lambda_H(m_t) = \lambda_H^{\rm exp}(m_t)$. 
The left panel of Figure~\ref{lambdaH2.FIG} shows the viable region for $\lambda_H$ as a function of $\tan\beta$ and $m_S$ with the other parameters being the same as in Figure~\ref{lambdaH1.FIG}.
The red line represents the central value of $\lambda_H^{\rm exp}$, and the darker and lighter pink regions indicate $1\sigma$ and $2\sigma$ values, respectively.
At small $\tan\beta$, $\lambda_H(\widetilde m)$ approaches zero, and thus $\lambda_H$ becomes negative at $\mu<\widetilde m$ since the $\beta$-function of $\lambda_H$ at $\mu=\widetilde m$ is positive.
Therefore, $\tan\beta \lesssim 1.23$ is disfavored in this case, which is shown as the gray shaded region labelled by $\lambda_H < 0$.
For the running of $\lambda_S$, since its $\beta$-function is always positive for $\mu=m_t$ to $\widetilde m$ in the parameter space of interest, it often becomes negative at low energy.
To prevent the negative $\lambda_S$ at $\mu=m_t$, the $S$ state should be decoupled before $\lambda_S(\mu)$ drops below zero, which is indicated by the gray shaded region labeled by $\lambda_S(m_t)<0$ in the figure.

The middle panel of Figure~\ref{lambdaH2.FIG} shows the viable region in the $m_S$-$\lambda(\widetilde m)$ plane.
The meaning of the gray region labeled by $\lambda_S(m_t)<0$ is the same as the left panel of the figure.
Similarly, $\lambda_{SS^*}$  also provides a constraint if it runs negative at low energy. Since $\lambda(\widetilde m)$ fixes the value of $\lambda_{SS^*}(\widetilde m)$, the latter can easily run negative at low $\mu$ for small $\lambda(\widetilde m)$.
This gives the limit indicated by the gray region labeled by $\lambda_{SS^*}(m_t)<0$.
Moreover, when $|\lambda(\widetilde m)| < \sqrt{3/34}g_2(\widetilde m)$, $\lambda_{SS^*}(\widetilde m)$ becomes negative, and thus this condition is considered as a lower bound on $\lambda(\widetilde m)$.
Therefore, although the running of $\lambda_H$ is not overly affected by $\lambda(\widetilde m)$, the viable parameter space is sensitive to this coupling since the running of $\lambda_S$ and $\lambda_{SS^*}$ strongly depends on it.
Note that from this figure, we see that our best fit value of $\lambda = 0.9082$ is excluded by these stability arguments.
However, we also see that from Figure \ref{chi2.FIG}, the value of $\chi^2 \approx 0.2$ when  we adjust $\lambda = 0.19$ and $\eta = 1$ (which has no effect on the running of the
quartic couplings)\footnote{We note that the comparison to Figure~\ref{chi2.FIG} is only approximate since $\lambda$ in that figure
should be evaluated at $\mu_\Phi$ whereas the value of $\lambda = 0.19$ to ensure vacuum stability is evaluated at ${\widetilde m}$. In addition, $\lambda(\mu_\Phi)  > \lambda (\widetilde m) = 0.19$.}. Thus gauge coupling unification for this choice of couplings remains perfectly acceptable. At these shifted values of $\lambda, \eta$, 
the proton lifetime exceeds the experimental limit, but is perhaps within the range of current experiments.

We next consider the RGE sensitivity to $\alpha(\widetilde m)$ and $\bar \alpha(\widetilde m)$ which has been implicitly assumed to be equal to $\alpha(\widetilde m)$.
The value of $\alpha(\widetilde m) = \bar \alpha(\widetilde m)$ determines the value of $\lambda_{HS}({\widetilde m})$.
In addition, the running of $\lambda_{HS}$ plays an important role in the running of $\lambda_H$, and hence $\lambda_H(m_t)$ is strongly dependent on
$\alpha(\widetilde m)$.
For large $\alpha(\widetilde m)$, $\lambda_H$ runs negative as seen in the right panel of Figure~\ref{lambdaH2.FIG} by the shaded region for large $\alpha$.  In this region, the contribution of $\lambda_{HS}$ to the $\beta$-function of $\lambda_H$ is too strong, and $\lambda_H$ quickly runs negative.
On the other hand, the effect of $\alpha(\widetilde m)$ in the running of $\lambda_S$ and $\lambda_{HS}$ is small, and $m_S$ can take a wide range of values.
Nevertheless, for small $m_S$, large $\alpha(\widetilde m)$ also causes $\lambda_S$ to run negative as seen by the lower shaded region.
Irrespective of stability criteria, we also see that obtaining the correct value for $\lambda_H(m_t)$ requires $\alpha(\widetilde m) = \bar \alpha(\widetilde m) = 0.57 - 0.64$.

\begin{figure}[ht!]
\centering
\includegraphics[scale=0.7]{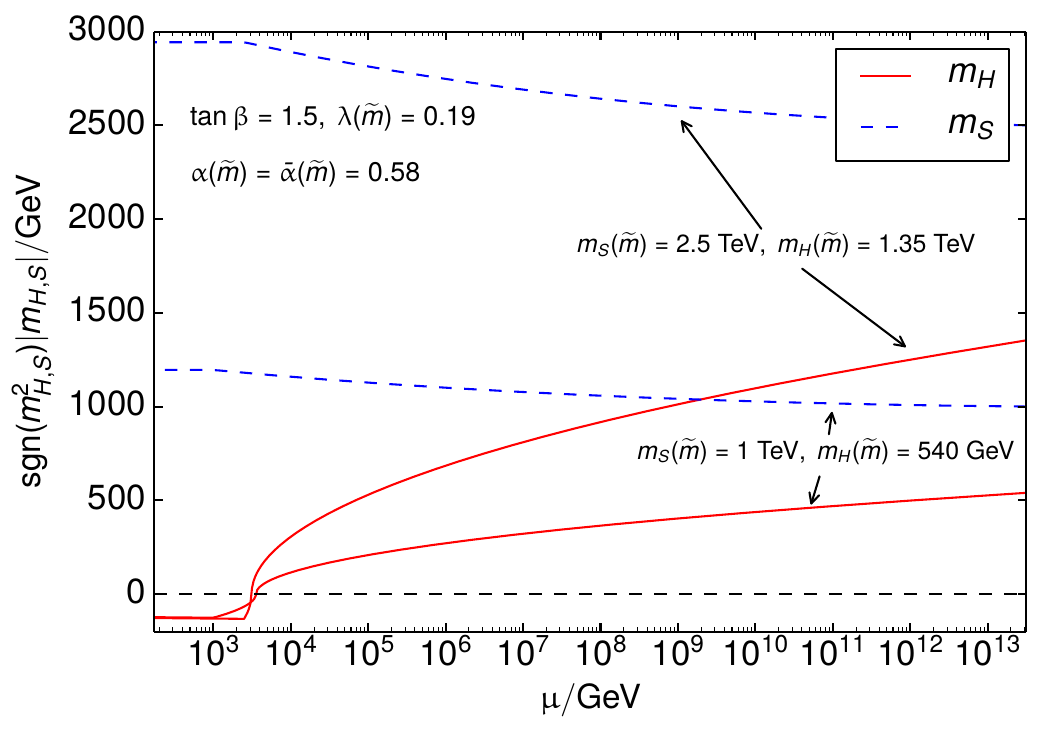}
\caption{The renormalization group evolution of ${\rm sgn}(m_H^2)|m_H|$ and ${\rm sgn}(m_S^2)|m_S|$ with $\tan\beta=1.5,~\lambda(\widetilde m)=0.19$ and $\alpha(\widetilde m)=\bar \alpha(\widetilde m)=0.58$. The black dashed line indicates the zero of the vertical axis. The Higgs mass-squared parameter $m_H^2$ runs negative at $\mu={\cal O}(10^3-10^4)~{\rm GeV}$, while $m_S^2$ remains positive.}
\label{ewsb1.FIG}
\end{figure}

\begin{figure}[ht!]
\centering
\includegraphics[scale=0.7]{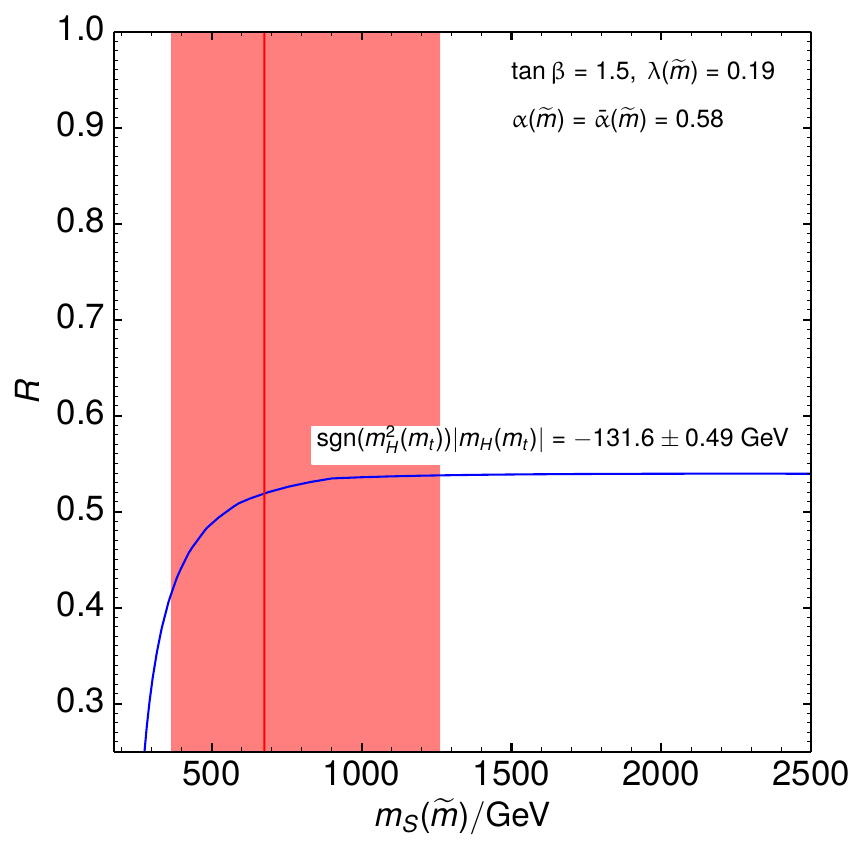}
\caption{A plot of the ratio $R$ as a function of $m_S({\widetilde m})$. The favored region for radiative electroweak symmetry breaking to occur is indicated by the blue solid line. The red solid line shows the required value of $m_S$ for the chosen value of $\lambda(\widetilde m)$ to obtain vacuum stability, and the pink region is its $1\sigma$ range.}
\label{ewsb2.FIG}
\end{figure}

Finally, we consider the running of the soft Higgs mass.
As stated in section 2, we have imposed the zero determinant condition, namely, ${\cal B}^{h\dagger}{{\cal B}^h}-|\widetilde m|^2\mathbb{1} = 0$, which fixes the value of $\mu_H$ to make $m_H(\widetilde m)=0$.
In practice, however, this condition can be relaxed to $m_H(\widetilde m)\neq 0$ as long as $m_H/\widetilde m\ll 1$, since the value of $\mu_H$ is only affected by an ${\cal O}(\mu_\Phi m_H^2/\widetilde m^2)$ term, and thus we take $m_H(\widetilde m)$ as a free parameter.
Figure~\ref{ewsb1.FIG} shows the running of $m_H$ and $m_S$, where the values of the relevant parameters are the same as in Figure~\ref{lambdaH1.FIG}, except for $m_S$.
The black dashed line in the figure indicates the zero of the vertical axis.
The red solid and blue dashed line represent the evolution of $m_H$ and $m_S$, respectively, and in the figure we take two different boundary values for $m_H(\widetilde m)$ and $m_S(\widetilde m)$.
In both cases, the electroweak symmetry is broken at $\mu= {\cal O}(10^3-10^4)$ GeV as $m_H$ runs negative, while $m_S$ maintains a positive value so that $S$ does not obtain a VEV.
For radiative EWSB to occur, the value of $m_H(\widetilde m)$ should be smaller than that of $m_S(\widetilde m)$, which is shown in Figure~\ref{ewsb2.FIG} where we define the ratio $R = m_H(\widetilde{m})/m_S(\widetilde{m})$.
The blue line shows the parameter region that satisfies the condition $\text{sgn} (m_H^2(m_t)) |m_H(m_t)|=-131.6\pm0.49$ GeV \cite{Buttazzo:2013uya}.
Above this line $m_H^2(m_t)$ becomes larger than the required value, and even remains positive.
The red solid line corresponds to the central value of $\lambda_H^{\rm exp}(m_t)$, and the pink shaded region shows the $1\sigma$ range.
The value of $m_S(\widetilde m)$ favored by vacuum stability can increase if we take a smaller value for $\tan\beta$, as expected from the left panel of Figure~\ref{lambdaH2.FIG}.
In any case, $R \lesssim |\alpha|$ should be satisfied for radiative EWSB to occur, as we will explain below.

If we consider the RGE of the Higgs potential mass-squared parameter, given here at one loop by
\begin{align}
(4\pi)^2 \beta^{(1)}_{m_H^2} = m_H^2\left( 6 y_t^2 + 12 \lambda_H^2 - \frac{9}{10}g_1^2 - \frac{9}{2}g_2^2\right) + 6 \lambda_{HS} m_S^2 \ , 
\label{mhsqRGE.EQ}
\end{align}
more closely, we can understand that radiative EWSB will always occur as long as the ratio $R$ is less than a specific value which is determined almost entirely by the superpotential parameters $\alpha$ and $\bar{\alpha}$. The RGE is dominated by the $y_t^2,~g_2^2$ and $\lambda_{HS} m_S^2$ terms, since $12 \lambda_H^2$ and $(9/10) g_1^2$ are small at all energy scales. For simplicity, as in our numerics, we shall take $\lambda_{HS} = |\alpha|^2$, which corresponds to choosing $|\alpha| = |\bar{\alpha}|$. Given this choice, we can rewrite the above $\beta$-function for the Higgs potential mass-squared parameter as 
\begin{align}
(4\pi)^2 \beta^{(1)}_{m_H^2} \simeq m_H^2\left( 6 y_t^2 - \frac{9}{2}g_2^2\right) + 6 |\alpha|^2 m_S^2 \ , 
\end{align}
which we can then use to solve the linearised RGE. Given the input parameters specified at the SUSY scale $\widetilde{m}$, the solution of the linearised RGE for the Higgs potential mass parameter at $m_S$ is 
\begin{align}
\nonumber m_H^2(m_S) \simeq m_H^2(\widetilde{m}) \Bigg\{ 1 - \frac{1}{(4\pi)^2}\Bigg[ &\left(6 y_t^2(\widetilde{m}) - \frac{9}{2}g_2^2(\widetilde{m})  + 6 |\alpha|^2 R^{-2}\right)\log\frac{\widetilde{m}}{m_S}  \Bigg] \Bigg\}\ ,
\end{align}
where we have replaced $m_S(\widetilde{m})$ by using $R = m_H(\widetilde{m})/m_S(\widetilde{m})$. Then we find that 
\begin{align}
m_H^2(m_S) \simeq - m_H^2(\widetilde{m}) \left( y_t^2(\widetilde{m}) - \frac{3}{4}g_2^2(\widetilde{m}) \right) \ ,
\end{align}
as long as the following condition for $R$ is satisfied:
\begin{align}
R^2 \simeq |\alpha|^2\frac{3 \log \frac{\widetilde{m}}{m_S}} {8\pi^2} \ .
\label{Rdef.EQ}
\end{align}
This quantity is quite close to being $|\alpha|^2$, since for the range of $m_S(\widetilde{m})$ which is phenomenologically viable, the fraction on the right of the above equation is $\sim 1$.

\begin{figure}[ht!]
\centering
\includegraphics[scale= 0.6]{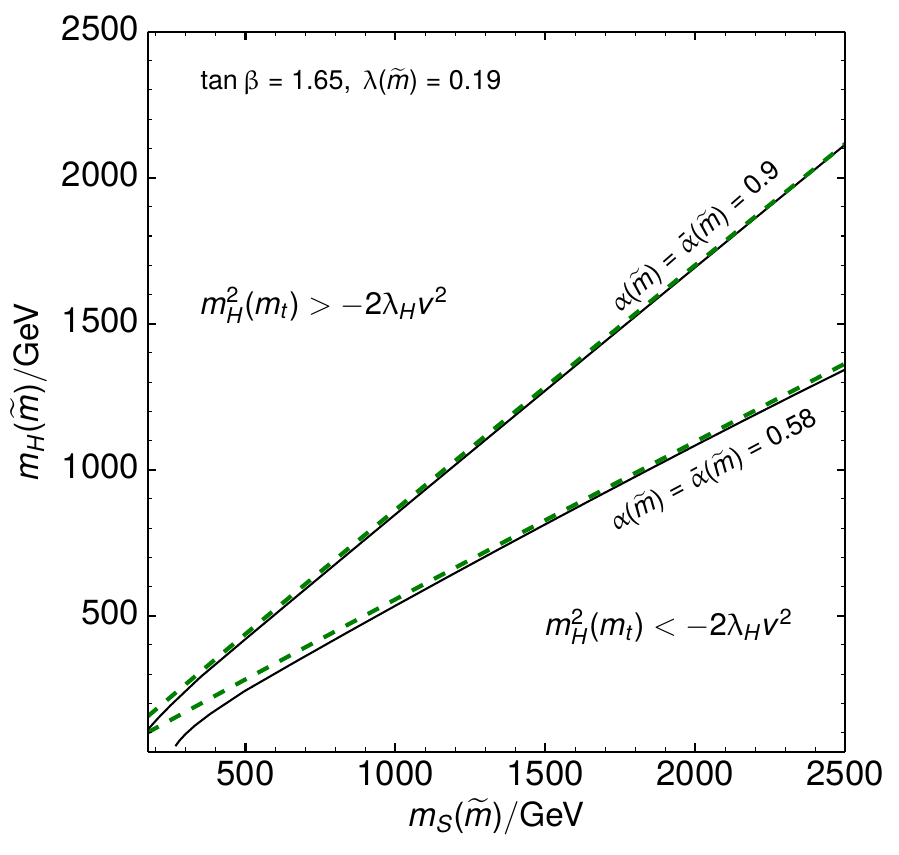}~~\includegraphics[scale= 0.6]{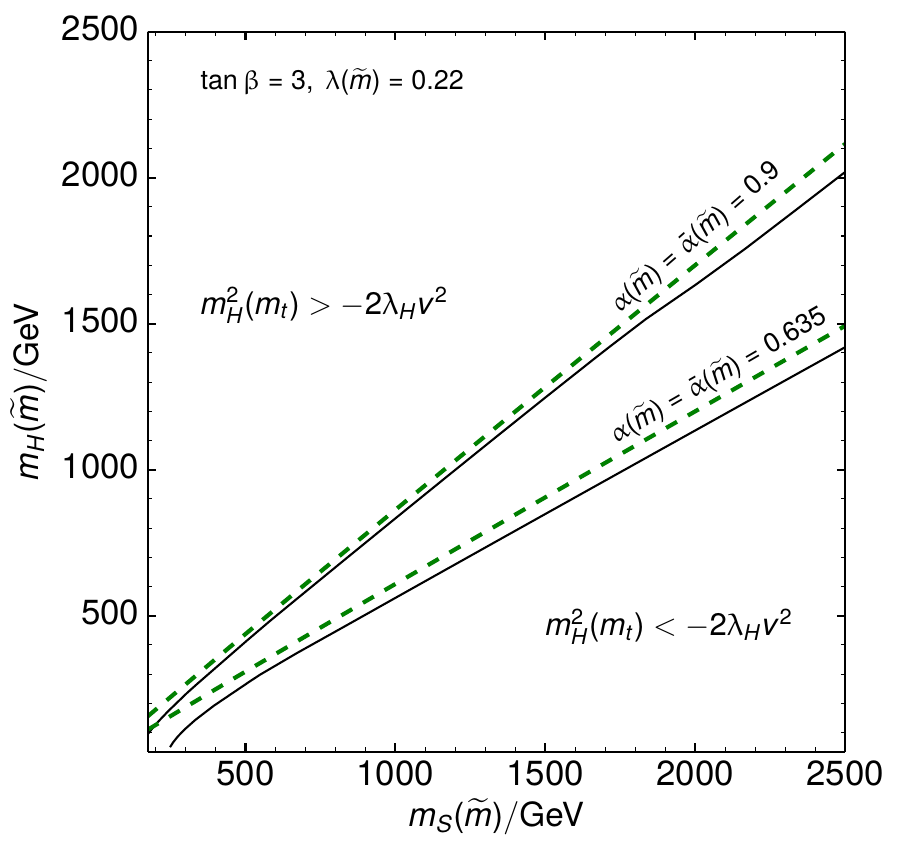}
\caption{Contours (black) where $m_H^2(m_t) = - 2\lambda_H v^2$, for chosen input values of $m_H(\widetilde{m})$ and $m_S(\widetilde{m})$. In dashed green are lines of constant $R$ defined by Eq. (\ref{Rdef.EQ}), corresponding to two choices of $\alpha$, showing how the estimated ratio corresponds quite closely to the numerical result, up to corrections depending on $\lambda(\widetilde{m})$. In the left panel we have chosen $\tan\beta=1.65$ and $\lambda(\widetilde{m})=0.19$ as a point corresponding to the lower end of the $m_S(\widetilde{m})$ preferred band in the left panel of Figure 10. In the right panel we have chosen $\tan\beta=3$ and $\lambda(\widetilde{m})=0.22$ to illustrate a point which does not correspond to the preferred band from Figure 10, but shows that up to small corrections, the semi-analytic estimate for $m_H^2(m_t)$ holds quite well. In both panels we have fixed $\widetilde{m} = 10^{13.5}$ GeV.}
\label{RPlot.FIG}
\end{figure}

We must then account for the running of the SM couplings properly to get $m_H^2(m_t)$. Since $y_t$ runs to large values in the IR, while $g_2$ does not run substantially, we find that as long as $R \lesssim |\alpha|$, $m_H^2(m_t) < 0$, and radiative EWSB is achieved. Indeed, as evidenced in Figure \ref{RPlot.FIG}, we find that as long as $R$ is given by Eq. (\ref{Rdef.EQ}), almost precisely the correct amount of symmetry breaking is achieved, with the low-scale boundary condition of 
\begin{align}
m_H^2(m_t) = - 2\lambda_Hv^2 \big|_{\mu=m_t}=-131.6~{\rm GeV} \ .
\end{align}
This result changes slightly for variations in $\lambda(\widetilde{m})$. This is due to the fact that the running of $\lambda_{HS}$ depends on $\lambda(\widetilde{m})$, and $m_S(\mu)$ depends on $\lambda(\widetilde{m})$ through $\lambda_{SS^*}$ directly, as well as through the running of $\lambda_{S}$. Thus, the SM domination in the relations above cannot be assumed to hold at all scales, and a full RGE analysis must be performed to find the exact value of $R$ which achieves the correct amount of radiative EWSB.

\section{Conclusions}

Despite the lack of discovery at the LHC \cite{nosusy}, weak-scale supersymmetry remains viable even in simplified models such as the constrained MSSM
(or its variants) \cite{eeloz}. Nevertheless, it is also quite possible that supersymmetry is manifest only at very high energies. 
If that is the case, it is important to consider whether any or all of the problems with cures normally attributed to weak-scale 
supersymmetry can still be resolved. With the exception of the hierarchy problem, we have argued that
gauge coupling unification can occur at the same time we ensure the stability of the Higgs vacuum, obtain radiative electroweak
symmetry breaking, and provide a dark matter candidate in a supersymmetric version of SO(10) when supersymmetry is broken 
above the inflationary scale, $3 \times 10^{13}$ GeV. In fact these issues can all be resolved in the context of non-supersymmetric
SO(10) \cite{moqz,mnoqz,noz, mnoz}. Because of the constraints on the vacuum structure imposed by supersymmetry
(even if broken at a high scale) solutions to these problems are not obvious.

While part of our initial motivation for high-scale supersymmetry was tied to gravitino dark matter
with a large reheating temperature \cite{DMO,dgmo,dgkmo}, our high-scale supersymmetric SO(10)
solution may provide a second candidate, namely the neutral component of a TeV scale scalar triplet, $S$,
which is a remnant of the {\bf 210} Higgs field breaking SO(10). While not expected to be stable,
its long lifetime may render it an acceptable (and perhaps detectable) dark matter candidate.

That some additional fields remain light is a necessary component of the model.
Unlike non-supersymmetric SO(10),  the supersymmetric version of SO(10)
does not have the luxury of choosing breaking patterns and intermediate scale
to ensure gauge coupling unification. The requirement of vanishing $F$- and $D$-terms
effectively breaks SO(10) directly to the SM with no intermediate scale to affect
the running of the gauge couplings. This means that some state must remain (be tuned to be) light.
As we have seen, although there are many possible representations within the {\bf 210}
or {\bf 126} and ${\bf \overline{126}}$, the only representation that 
achieves satisfactory gauge coupling unification is the (1,3,0) component of the {\bf 210},
our weak scalar triplet $S$. 

It is also clear that to resolve the problem of the stability of the Higgs vacuum in the SM,
some state must remain light (at least below $10^{10}$ GeV) in order to deflect the running 
of the Higgs quartic coupling so that it remains positive as it runs towards the ultra-violet. 
Thus our $S$ state serves to assist in gauge coupling unification, protect the Higgs vacuum and 
due to its long lifetime, perhaps provide a dark matter candidate. This may explain why a second 
tuning beyond having a light Higgs is needed.

A resounding issue surrounding high-scale supersymmetry is verifiability. 
The sparticle spectrum is all assumed to be so heavy that that
sparticles were never part of the thermal background after inflationary 
reheating. The only $R$-parity odd state below the inflationary scale is 
a gravitino with mass in excess of 0.1 EeV \cite{DMO} and thus one would
expect that accelerator and direct detection searches would come up empty.
If $R$-parity is violated, a long lived gravitino may provide an indirect signal through
very high energy monochromatic neutrinos \cite{dgkmo}. 
In the current model, we have shown that retaining most of the advantages of weak-scale
supersymmetry in high-scale supersymmetry requires a weak-scale state SU(2)$_L$ triplet scalar $S$. Therefore, this state may provide a window into high-scale supersymmetry, 
and its cosmology and phenomenology will be studied more fully in future work.

\section*{Acknowledgments}
S.A.R.E. is supported by the SNF through grant P2SKP2 171767 and by the U.S. Department of Energy under Contract No. DE-AC02-76SF00515. The work of T.G., K.K., and  K.A.O. was supported in part by the DOE grant DE--SC0011842 at the University of Minnesota.

\end{document}